\documentclass{aastex6}
\usepackage{amsmath}
\usepackage[normalem]{ulem}
\usepackage{appendix}
\usepackage{graphicx}
\usepackage{array,tabularx}

\def\numax{$\nu_{\rm max}$}

\def\dnu{$\Delta\nu$}
\def\teff{$T_{\rm eff}$}
\def\logg{$\log g$}
\def\be{\begin{equation}}
\def\ee{\end{equation}}
\def\FeH{$\mathrm{[Fe/H]}$}
\def\Alpha{$\alpha$}

\defcitealias{Magic2015}{M15}
\defcitealias{Trampedach2011}{T11}
\defcitealias{Trampedach2013}{T13}
\defcitealias{Trampedach2014}{T14}

\begin{document}

\title{Investigating the Metallicity--Mixing Length Relation}

\author{Lucas S. Viani\altaffilmark{1}, Sarbani Basu\altaffilmark{1}, Joel Ong J. M.\altaffilmark{1}, Ana Bonaca\altaffilmark{2}, and William J. Chaplin\altaffilmark{3,4}}
 
\altaffiltext{1}{Department of Astronomy, Yale University, New Haven, CT, 06520, USA}

\altaffiltext{2}{Department of Astronomy, Harvard University, Cambridge, MA, 02138, USA}

\altaffiltext{3}{School of Physics and Astronomy, University of Birmingham, Edgbaston, Birmingham, B15 2TT, UK}

\altaffiltext{4}{Stellar Astrophysics Centre (SAC), Department of Physics and Astronomy, Aarhus University, Ny Munkegade 120, DK-8000 Aarhus C, Denmark}

\email{lucas.viani@yale.edu}

\begin{abstract}
Stellar models typically use the mixing length approximation as a way to implement convection in a simplified manner. While conventionally the value of the mixing length parameter, \Alpha, used is the solar calibrated value, many studies have shown that other values of \Alpha\ are needed to properly model stars. This uncertainty in the value of the mixing length parameter is a major source of error in stellar models and isochrones. Using asteroseismic data, we determine the value of the mixing length parameter required to properly model a set of about 450 stars ranging in \logg, \teff, and \FeH. The relationship between the value of \Alpha\ required and the properties of the star is then investigated. For Eddington atmosphere, non-diffusion models, we find that the value of \Alpha\ can be approximated by a linear model, in the form of $\alpha/\alpha_{\sun}=5.426 -0.101 \log (g)   -1.071 \log (T_{\mathrm{eff}}) + 0.437 (\mathrm{[Fe/H]})$. This process is repeated using a variety of model physics as well as compared to previous studies and results from 3D convective simulations.
\end{abstract}


\section{Introduction}
\label{sec:Intro}
One of the largest issues in producing accurate stellar models is handling the complexities of stellar convection. The convective process is typically approximated by implementing the ``mixing length theory'' or MLT \citep{Bohm1958}. Here, convective eddies are assumed to have an average size of $\alpha H_p$ where $H_p$ is the pressure scale height and \Alpha, a free parameter in the models, is referred to as the mixing length parameter. It is also assumed the the convective eddies on average can travel a distance of $\alpha H_p$ before they lose their identity. Therefore, \Alpha\ determines the efficiency of convection. In low mass stars, \Alpha\ also determines the radius.

Since the value of the mixing length parameter does not have a physical basis, the process determining the value of \Alpha\ that should be used to model a star is not obvious. Typically, for a given set of model physics, a calibration is performed to determine what value of \Alpha\ is needed to reproduce the global properties of the Sun. In other words, what value of \Alpha\ will give a 1 $R_{\sun}$ and 1 $L_{\sun}$ star at the solar age. This solar calibrated \Alpha\ is then used as the value of the mixing length parameter for the other stars using this same set of model physics. While this is the typical process, it is not clear that stars with different properties should all have the same value of \Alpha. Indeed, it is now known that assuming all stars should have the same \Alpha\ as the solar calibrated value is incorrect. For example, \cite{Lattanzio1984} and \cite{Demarque1986} demonstrated that if the solar calibrated mixing length is used then the radius of $\alpha$ Cen A cannot be accurately modeled. Similarly, \cite{GuentherDemarque2000} found that $\alpha$ Cen A and B should have different mixing length values which are both also different than the solar calibrated value, if it is assumed that the two stars are of the same age and composition. \cite{Fernandes1995}, \cite{Eggenberger2004}, and \cite{Miglio2005} also found differing \Alpha\ values for the two stars as well. For 16 Cyg A \& B \cite{Metcalfe2012} found that \Alpha\ needed to be different than the solar value. Similarly, for Procyon A \cite{Straka2005} concluded that a value of \Alpha\ different from solar was needed. \cite{Joyce2017} demonstrated that subsolar values of \Alpha\ were needed to model globular cluster M92 as well as 5 low metallicity ($\mathrm{[Fe/H]}\sim-2.3$) stars by fitting stellar models to observed non-seismic properties ($T_{\mathrm{eff}}$, $L$, $R$, and $\mathrm{[Fe/H]}$). Also, \cite{Chun2018} show that in red supergiants the mixing length increases as metallicity increases. Additionally, \cite{DeheuvelsMichel2011}, \cite{Mathur2012}, \cite{Metcalfe2014}, and many other studies have showed that different (non-solar) values of \Alpha\ are needed to properly model different stars.

These findings showing the errors in simply assuming a solar calibrated mixing length parameter inspired further investigation. \cite{Bonaca2012} found that to accurately model stars with data from NASA's \textit{Kepler} mission, the value of \Alpha\ needed to be lower than the solar calibrated mixing length. They also showed that using the solar calibrated value of \Alpha\ often resulted in stars having initial helium abundances lower than the primordial helium abundance. Additionally, \cite{Bonaca2012} examined possible trends between \Alpha\ and stellar properties, finding that \Alpha\ increased with metallicity. Using convection simulations, \cite{Tanner2014} also found a relation between \Alpha\ and metallicity. Studies of binary system have shown that \Alpha\ could be linked to the mass of a star \citep[e.g.,][etc.]{LudwigSalaris1999, Morel2000, Lebreton2001, Lastennet2003, Yildiz2006}. This relationship between \Alpha\ and mass may be explained by the findings of an \Alpha\ dependence on \teff\ and \logg\ \citep[e.g.,][etc.]{LudwigFreytagSteffen1999,Trampedach2007,Trampedach2011,Trampedach2013,Tanner2013,Magic2013}.

This study will expand upon the sample of stars used in \cite{Bonaca2012}, covering a larger parameter space in \logg, \teff, and \FeH. The metallicity values used in this work \citep[from][]{Buchhave2015} are also more accurate. Additionally, the method by which the model values are fit to the observations is improved. In \cite{Bonaca2012} the observed values of \dnu, \numax, \teff, and \FeH\ for each star were input into the grid-based Yale-Birmingham pipeline \citep{Basu2010,Basu2012,Gai2011} to give estimates of mass and radius. The created stellar models were then fit to this mass and radius to determine the value of \Alpha. However, at different \teff\ and \FeH\ values there is no guarantee that these models were good fits to the original asteroseismic values of \dnu\ and \numax. In this work, we will instead fit directly to the asteroseismic properties of the star. Additionally, we will not rely on the scaling relations, instead calculating the value of \dnu\ for the model stars using their radial-mode frequencies.

The paper is organized as follows: Section~\ref{sec:models_and_analysis} explains the details of the stellar models, the handling of the surface term correction, and the model likelihood calculations. Section~\ref{sec:results} provides the results for the base set of models as well as the other sets of model physics. Section~\ref{sec:Discussion} compares these results to studies from 3D convection simulations as well as other studies and presents the conclusions.

\section{Data, Models, and Analysis}
\label{sec:models_and_analysis}

\subsection{Data}
\label{sec:Data}

Each star in this study has observed measurements of $\nu_{\mathrm{max}}$, $\Delta \nu$, $T_{\mathrm{eff}}$, and $\mathrm{[Fe/H]}$. $\nu_{\mathrm{max}}$ is the frequency at which the oscillation power is at a maximum and can be approximately related to a star's surface gravity and effective temperature as $\nu_{\mathrm{max}} \propto g T_{\mathrm{eff}}^{-1/2}$ \citep{Brown1991,Kjeldsen1995,Bedding2003}. $\Delta \nu$, the large frequency separation, is the average frequency spacing between adjacent radial order ($n$) modes of the same degree ($\ell$). $\Delta \nu$ is approximately related to a star's average density by $\Delta \nu \propto \sqrt{\bar{\rho}}$ \citep[see, e.g.,][]{Tassoul1980,Ulrich1986,Dalsgaard1988,Dalsgaard1993}. The seismic data was obtained from \cite{Serenelli2017}, who added 415 subgiant and dwarf stars to the original APOKASC catalog \citep{Pinsonneault2014} and determined $\Delta \nu$ and $\nu_{\mathrm{max}}$ values from \textit{Kepler} lightcurves. The \FeH\ values were obtained from the spectroscopic survey of \cite{Buchhave2015}. \cite{Buchhave2015} observed the stars with the Tillinghast Reflector Echelle Spectrograph using the 1.5 m Tillinghast Reflector. The $T_{\mathrm{eff}}$ values were determined from these spectra in an iterative process after fixing $\log g$ to seismic values and can be found in \cite{MathurTemps}. These stars and their properties can be seen in Table~\ref{table:star_data}.

\begin{figure*}[h]
\epsscale{1.0}
\plotone{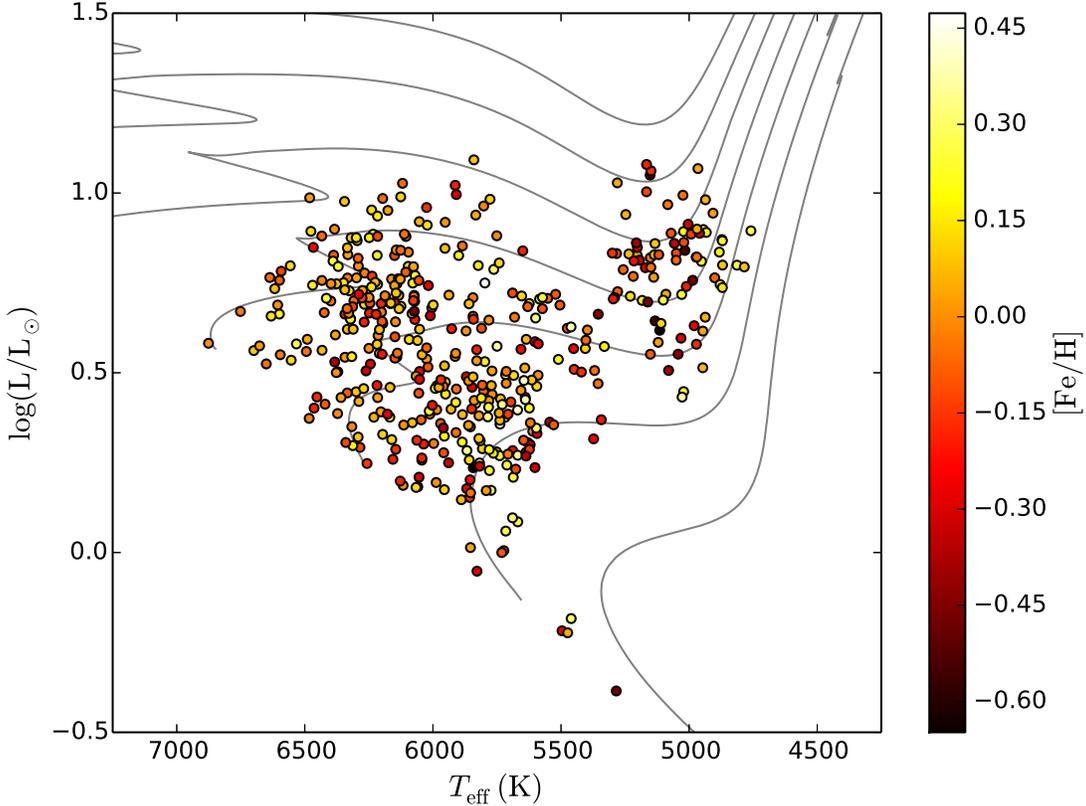}
\caption{H-R diagram for the stars in this study. The stellar properties were obtained after the modeling process was completed. The points are colored by their likelihood weighted average value of \FeH. The background gray lines are tracks of 0.8, 1.0, 1.2, 1.4, 1.6, 1.8, and 2.0 $M_{\sun}$ generated using YREC \citep{Demarque2008}.}
\label{fig:HR_plot}
\end{figure*}

Figure~\ref{fig:HR_plot} shows the H-R diagram with the stars in this work included, for the non-diffusion models. The background gray lines show tracks of 0.8, 1.0, 1.2, 1.4, 1.6, 1.8, and 2.0 $M_{\sun}$, generated using YREC \citep{Demarque2008}, for reference. The values of $T_{\mathrm{eff}}$ and $L$ are from the likelihood weighted average values of the models (as explained in Sec.~\ref{sec:likelihoods}). The points are colored by their likelihood weighted average value of \FeH.

\subsection{Constructing the Models}
\label{sec:models}

Each star in our sample was modeled using the Yale stellar evolution code YREC \citep{Demarque2008}. All models were created with the OPAL equation of state \citep{Rogers2002} and OPAL opacities \citep{Iglesias1996} supplemented with low temperature opacities from \cite{Ferguson2005}. Nuclear reaction rates from \cite{Adelberger1998} were adopted, except for that of the $^{14}N$($p$,$\gamma$)$^{15}O$ reaction, for which we used the rates of \cite{Formicola2004}. Models were constructed with Eddington gray atmospheres.

The core set of models include core overshoot with an extent of $0.2H_p$. We did construct a subset of models without core overshoot as well. We made two full sets of models, the first of which did not include the diffusion and gravitational settling of helium and other heavy elements. These ``No Diffusion'' models form the core of our investigation. The second set of models were constructed including diffusion and gravitational settling using the rates of \cite{Thoul1994}. However, in hot stars, diffusion as modeled is known to drain out heavy elements quickly and to avoid this we multiplied the diffusion rate by a mass-dependent factor given by
\begin{equation}
\begin{cases} 
  \mathcal{F}_{\mathrm{Diffusion}}=\exp\left(   \frac{-(M-1.25)^2}{2*0.085^2}   \right) , & M > 1.25 M_{\sun} \\
  \mathcal{F}_{\mathrm{Diffusion}}=1, & M \leq 1.25 M_{\sun} 
\end{cases}
\label{eq:diffusion_coef}
\end{equation}
Eq.~\ref{eq:diffusion_coef} serves to smoothly decrease the diffusion rate for higher mass stars. The need to use this rather arbitrary factor is why we use the non-diffusion models as our primary set.

The $\Delta Y/\Delta Z$ relation was determined by constructing standard solar models. For models without diffusion, a calibrated solar model implied $Y=0.248+1.0958Z$, where we assumed that the primordial helium abundance $Y_p=0.248$. For models with diffusion, we get $Y=0.248+1.4657Z$. We use the metallicity scale of \cite{Grevesse1998} to convert [Fe/H] to $Z/X$. Note that the solar calibrated mixing length is 1.70098 for models without diffusion and 1.838417 for models with diffusion.

The starting point of modeling each star was the input \dnu, \numax, \teff\ and [Fe/H]. Since each of these quantities is associated with an uncertainty, we created many more realizations of these parameters to obtain a larger set of (\dnu, \numax, \teff, [Fe/H]), however, in each case the uncertainty was assumed to be 1.5 times larger than the quoted uncertainties in the data to obtain a larger range of inputs, this was particularly important in order to ensure that surface term effects on \dnu\ and \numax\ do not bias the calculations at this stage. Each realization of the inputs was then used to calculate mass $M$ and radius $R$ of the models using the modified \dnu\ relation proposed by \cite{Guggenberger2016} and the usual \numax\ scaling relation \citep{Kjeldsen1995}. Note that the scaling relations here are used simply to determine a model mass and radius for each realization. When actually analyzing the models and determining model likelihoods (see Sec.~\ref{sec:likelihoods}) the scaling relations are not used, as individual model frequencies are calculated instead, as will be described shortly. Models were constructed for each ($M$, $R$, \teff, [Fe/H]) realization. This was done by running YREC in an iterative manner by allowing the the mixing length parameter $\alpha$ to vary until we constructed a model of the required radius at the required \teff\ for the given mass and metallicity. 

We calculated the $\ell=0$ mode frequencies for each model. These were then used to determine the large separation \dnu\ as an average of the large separation weighted by the observed power envelope which is a Gaussian with a FWHM of $0.66\nu_{\rm max}^{0.88}$ \citep{Mosser2012}. The obtained \dnu\ was corrected for surface-term effects using the factor determined in Section~\ref{sec:surface_correction}. For each model \numax\ was calculated from the acoustic cutoff using the prescription of \cite{Viani2017}. The surface-term corrected \dnu, \numax, \teff\ and [Fe/H] were then used to calculate the likelihood for each model (Section~\ref{sec:likelihoods}).

We used a minimum of 500 realizations, though most stars needed more. The number of realizations was determined by determining $\alpha$ after every ten runs beyond the initial 500, and determining if the likelihood weighted average of $\alpha$ (see Section~\ref{sec:likelihoods}) converged. Note that our procedure was different from that of \cite{Bonaca2012}, who first determined $M$ and $R$ from a grid-based modeling exercise and then did a Monte Carlo over the derived ($M$, $R$, \teff, [Fe/H]). We realized that the old procedure could lead to models that do not satisfy the input \dnu\ and \numax.

\subsection{Surface Term Corrections}
\label{sec:surface_correction}
When using frequencies of stellar models, the so-called ``surface term'' \citep{CDandBerthomieu1991}, a frequency difference between observed frequencies and frequencies from stellar models, must be considered and corrected. The surface term has been found to be a function of frequency once corrected for the mode-inertia \citep{CDandBerthomieu1991}. These frequency differences arise due to the difficulties in modeling convection and the upper layers of stars. This therefore means that the surface term is also model dependent and is affected by the model physics \citep{CDandThompson1997,Dziembowski1988}. Many different methods of correcting for the surface term exist, for example modeling the surface term as a power law, using a scaled version of the solar surface term, as well as more complicated methods \citep[e.g., see][]{Kjeldsen2008, Gruberbauer2012, Ball2014, Silva2015}. An extensive comparison of these methods are discussed in more detail in \cite{Schmitt2015}. For individual frequency modes, \cite{Ball2014} show that the frequency shift can be approximated well with the function form,
\begin{equation}
\delta \nu = \frac{1}{\mathcal{I}} \left[ a_{-1} \left( \frac{\nu}{\nu_{\mathrm{ac}}}\right)^{-1} + a_{3} \left(\frac{\nu}{\nu_{\mathrm{ac}}}\right)^{3}\right],
\label{eq:Ball_correction}
\end{equation}
where $\delta \nu$ is the frequency shift, $\mathcal{I}$ is the normalized mode inertia, $\nu_{\mathrm{ac}}$ is the acoustic cutoff frequency, and $a_{-1}$ and $a_{3}$ are the coefficient variables to be fit.

Since one of the stellar parameters these models will be fit to is $\Delta \nu$, some type of surface term correction must be applied in order to get correct results. In other words, we must determine how the surface correction effects our model measurements of $\Delta \nu$. A comparison of the value of $\Delta \nu$ determined from stellar models and from observations was performed for two different data sets. The first data set, from \cite{Lund2017}, consisted of observed mode frequencies from the \textit{Kepler} LEGACY sample. These \textit{Kepler} LEGACY sample stars had also previously been modeled using YREC for use in \cite{Silva2017}. For these models the individual mode frequencies were determined using the ``Yale Monte Carlo Method'' (YMCM) described in \cite{Silva2015} with some slight modifications as explained in \cite{Silva2017}. Models for these \textit{Kepler} LEGACY sample stars were created both with and without diffusion. The second data set has observed oscillation modes from \textit{Kepler} from \cite{Davies2016}. Model frequencies were obtained using the YMCM with models from \cite{Silva2015}. For this set, only stars with measured $\nu_{\mathrm{max}}$ values were included.

For each stellar model, the \cite{Ball2014} surface term correction (Eq.~\ref{eq:Ball_correction}) could be applied. Since the individual mode frequencies were computed for each model and the corresponding observed frequencies were known from the literature sources mentioned, the coefficients $a_{-1}$ and $a_{3}$ were solved for each star by implementing a two-term unweighted least-$\chi^2$ fit. The surface term corrected frequencies for the stellar models were then compared to the observed frequencies to determine which stellar models were the best for each star.  

\begin{figure*}[h]
\epsscale{1.1}
\plotone{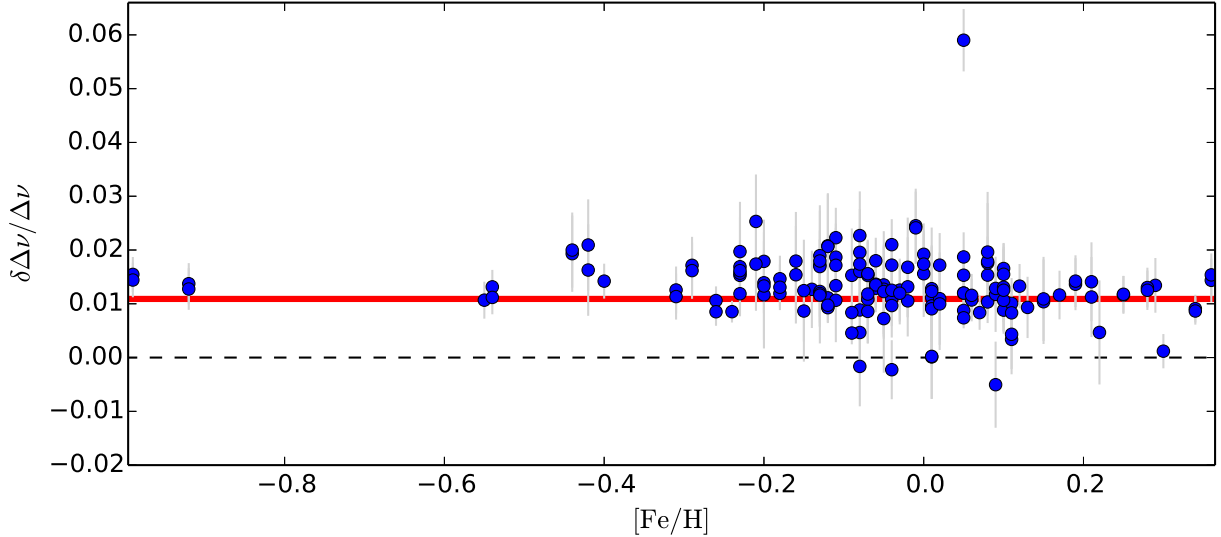}
\caption{The fractional difference, $\delta (\Delta \nu) / \Delta \nu$, between the models and observations. The red line shows the error weighted average value of $\delta (\Delta \nu) / \Delta \nu$ while the black dashed line at 0 is for reference.}
\label{fig:Joel_2}
\end{figure*}

Then, for each set of models $\Delta \nu$ was calculated for each star. For the model stars the large frequency separation was calculated by taking the individual mode frequencies for the $\ell=0$ modes and the corresponding radial quantum number ($n$) and using a weighted least-squared linear fit with Gaussian weights centered around $\nu_{\mathrm{max}}$ with a FWHM of $0.66 \nu_{\mathrm{max}}^{0.88}$ as described in \cite{Mosser2012}. The fractional difference between the model calculated value of $\Delta \nu$ and the observational value of $\Delta \nu$ for each model was then calculated where $\delta (\Delta \nu)/\Delta \nu = (\Delta \nu_{\mathrm{model}} - \Delta \nu_{\mathrm{obs}}) / \Delta \nu_{\mathrm{obs}}$. This fractional difference between $\Delta \nu$ from the models and from the observed data can be seen in Fig.~{\ref{fig:Joel_2}}.

From the values of $\delta (\Delta \nu)/\Delta \nu$ we can estimate the factor by which our model value of $\Delta \nu$ differs from the observed value of $\Delta \nu$. Based on the data in Fig.~{\ref{fig:Joel_2}}, for the error-weighted average for all the models together we obtain $\delta (\Delta \nu)/\Delta \nu = 0.0109$. This means that 
\begin{equation}
\Delta \nu_{\mathrm{obs}} \approx \Delta \nu_{\mathrm{model}} / 1.0109,
\label{eq:Eddington_surface_correction}
\end{equation} 
and so using Eq.~{\ref{eq:Eddington_surface_correction}} the surface term effects on $\Delta \nu$ can be removed and we can have confidence in comparing our model $\Delta \nu$ values to the observed $\Delta \nu$ values of our sample stars.

\subsection{Model Likelihoods and Determining Stellar Quantities}
\label{sec:likelihoods}
For each star, the Monte Carlo simulations resulted in typically 2-3 thousand stellar models. For each model, likelihood values for \dnu, \numax, \teff, and \FeH, were computed. These likelihood values were calculated following the form of Eq.~\ref{eq:likelihoods}, using \dnu\ as an example.

\begin{equation}
\mathcal{L}_{\Delta \nu} = \frac{1}{ \sqrt{2 \pi}  \sigma_{\Delta \nu}} \exp\left( \frac{-(\Delta \nu_{\mathrm{obs}}-\Delta \nu_{\mathrm{model}})^2}{2 \sigma_{\Delta \nu}^2}\right) 
\label{eq:likelihoods}
\end{equation}
where $\Delta \nu_{\mathrm{model}}$ is the model's value of $\Delta \nu$, $\Delta \nu_{\mathrm{obs}}$ is the observed value of $\Delta \nu$ from \cite{Serenelli2017}, and $\sigma_{\mathrm{\Delta \nu}}$ is the uncertainty in the observed value of $\Delta \nu$. Likelihood values for \dnu, \numax, \teff, and \FeH, were calculated in this manner. Note that before calculating the value of $\mathcal{L}_{\Delta \nu}$ the model values of $\Delta \nu$ were corrected for the surface term as explained in Sec.~{\ref{sec:surface_correction}}. 

\begin{figure}[h]
\begin{center}
\includegraphics[width=2 in]{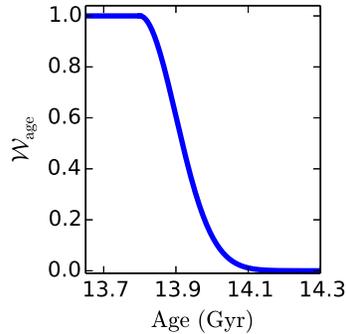}
\end{center}
\caption{The value of $\mathcal{W}_{\mathrm{age}}$ (Eq.~\ref{eq:likelihoods_Age}) for a range of ages.}
\label{fig:age_weights}
\end{figure}

A weighting factor for the age of the models was also included, with the purpose to ensure that models older than the age of the universe were given lower weights. The age weighting factor, $\mathcal{W}_{\mathrm{age}}$, is given by
\begin{equation}
\begin{cases} 
  \mathcal{W}_{\mathrm{age}} = \exp\left( \frac{-(\mathrm{Age}_{\mathrm{Universe}}-\mathrm{Age}_{\mathrm{model}})^2}{2 \sigma_{\mathrm{Age_{\mathrm{Universe}}}}^2}\right), & \mathrm{Age}_{\mathrm{model}} > \mathrm{Age}_{\mathrm{Universe}} \\
  \mathcal{W}_{\mathrm{age}} = 1, & \mathrm{Age}_{\mathrm{model}} \leq \mathrm{Age}_{\mathrm{Universe}} 
\end{cases}
\label{eq:likelihoods_Age}
\end{equation}
where the age of the universe is 13.8 Gyr and $\sigma_{\mathrm{Age_{\mathrm{Universe}}}}=0.1$ Gyr. The value of $\sigma_{\mathrm{Age_{\mathrm{Universe}}}}$ was chosen such that the weighting function smoothly and quickly goes to zero for high ages (as can be seen in Fig.~\ref{fig:age_weights}). 

The total likelihood for each model was determined by multiplying the individual likelihood values together, so that
\begin{equation}
\mathcal{L} = \mathcal{L}_{\Delta \nu} \: \mathcal{L}_{\nu_\mathrm{max}} \: \mathcal{L}_{T_\mathrm{eff}} \: \mathcal{L}_{\mathrm{[Fe/H]}} \: \mathcal{W}_{\mathrm{age}}   
\label{eq:likelihoods_total}
\end{equation} 
With the likelihood of each model for a given star determined, the likelihood weighted average of model quantities was calculated following the form of Eq.~\ref{eq:likelihoods_average}, using mass as an example,
\begin{equation}
\langle M \rangle = \frac{\sum_{i}^{N} M_i \mathcal{L}_i}{\sum_{i}^{N} \mathcal{L}_i}
\label{eq:likelihoods_average}
\end{equation}
with the likelihood weighted uncertainty being
\begin{equation}
\sigma^2 = \frac{\sum_{i}^{N} (M_i-\langle M \rangle)^2 \mathcal{L}_i }{\sum_{i}^{N} \mathcal{L}_i},
\label{eq:likelihoods_uncert}
\end{equation}
where $N$ is the number of models for each star. Using Eq.~\ref{eq:likelihoods_average} and \ref{eq:likelihoods_uncert} the value of stellar properties, such as mass, radius, age, temperature, \FeH, and \Alpha\ were calculated for each star. The process was also recalculated, for later use (see Sec.~\ref{sec:other_physics}), excluding the \numax\ likelihood values in Eq.~\ref{eq:likelihoods_total} so that the total likelihood was only determined by the \dnu, \teff, \FeH, and age values. 

\section{Results}
\label{sec:results}

\subsection{Probability Density Functions}
\begin{figure}[h]
\begin{center}
\includegraphics[width=3 in]{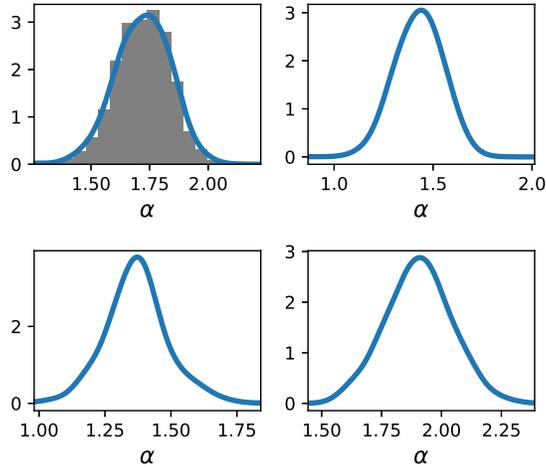}
\end{center}
\caption{Examples of symmetric PDFs for the mixing length parameter. Clockwise from top left the example stars are KIC 3329196, 4038445, 4914923, and 5429911. The gray histogram in the first panel shows the discrete PDF with the blue lines showing the continuous and smoothed PDF, generated using a KDE.}
\label{fig:normal_PDFs}
\end{figure}

\begin{figure}[h]
\begin{center}
\includegraphics[width=3.0 in]{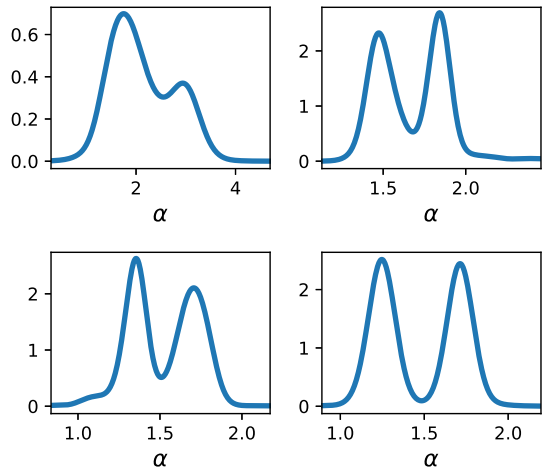}
\end{center}
\caption{Examples of bimodal PDFs for the mixing length parameter. Clockwise from top left the example stars are KIC 1430163, 3223000, 3967859, and 12265063.}
\label{fig:bimodal_PDFs}
\end{figure}

\begin{figure}[h]
\begin{center}
\includegraphics[width=3.0 in]{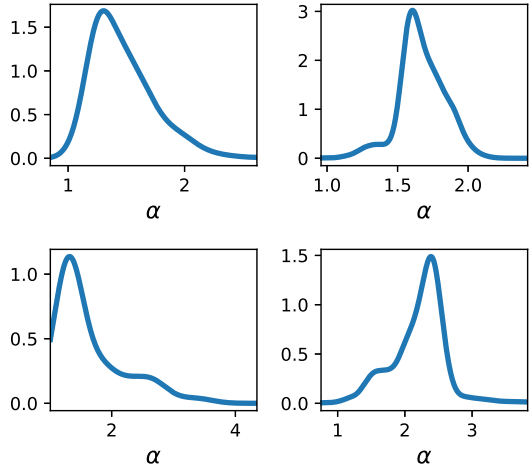}
\end{center}
\caption{Examples of asymmetric PDFs for the mixing length parameter. Clockwise from top left the example stars are KIC 6129877, 1725815, 3661135, and 9328372.}
\label{fig:skewed_PDFs}
\end{figure}

\begin{figure}[h]
\begin{center}
\includegraphics[width=3.0 in]{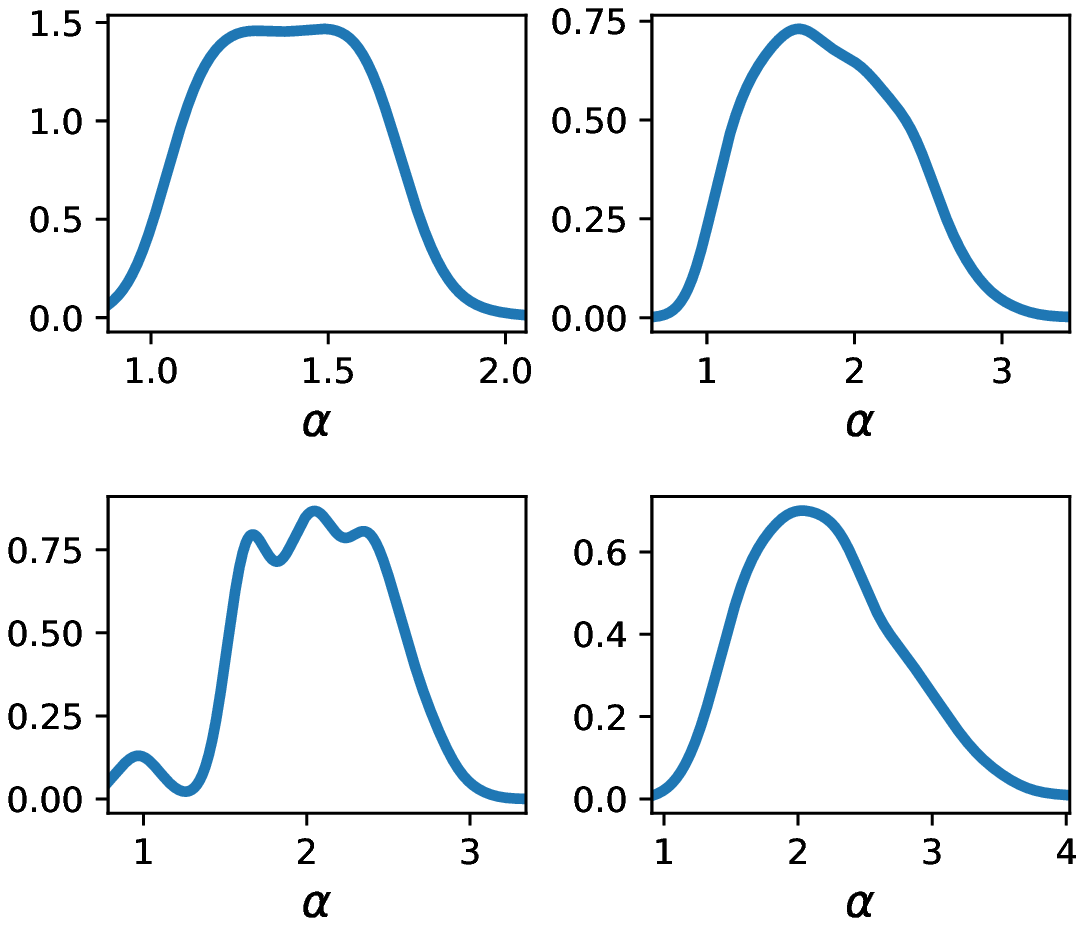}
\end{center}
\caption{Examples of wide PDFs for the mixing length parameter. Clockwise from top left the example stars are KIC 3657002, 5543462, 8172589, and 9005973.}
\label{fig:wide_PDFs}
\end{figure}

For each star the probability density function (PDF) for \Alpha, that was obtained by marginalizing over all other parameters, was also examined to ensure the models converged properly. Several examples of the different types of PDFs (symetric, bimodal, skewed, and wide) for \Alpha\ can be seen in Figures~\ref{fig:normal_PDFs}-\ref{fig:wide_PDFs}. These figures show the smoothed continuous PDF functions. An example of the discrete PDF can be seen in the first panel of Fig.~\ref{fig:normal_PDFs} (the gray background histogram). The discrete PDFs were transformed into the smoothed continuous PDFs using a kernel density estimation (KDE), where the bandwidth was determined using Scott's rule \citep{Scott1992}. Note that the stellar properties were not determined from the PDFs but from the likelihood weighted averages.

\subsection{Examining the Relationship between $\alpha$ and Metallicity}
Following the work of \cite{Bonaca2012}, we are interested in examining the relationship between the mixing length parameter and a star's metallicity. First we will examine the non-diffusion, $0.2H_p$ set of stellar models and in later sections we report the result of the analysis for the other sets of models. Since we will later be comparing the results to models with diffusion, it is best to use the quantity $\alpha / \alpha_{\sun}$ to compare the results of the different model physics. As mentioned earlier, in the non-diffusion case the value of $\alpha_{\sun}$ is $1.70098$ and for the diffusion models $\alpha_{\sun}$ is $1.838417$. The analysis is performed on stars with likelihood weighted \Alpha\ of $<4$, as the results for stars with $\alpha > 4$ are rather unstable and thus not included. For the 460 stars modeled in this work, only 7 had average likelihood weighted values of $\alpha>4$. The high $\alpha$ values for the handful of excluded stars arose due to the models not properly converging in those cases.
 
\begin{figure*}[h]
\epsscale{1.1}
\plotone{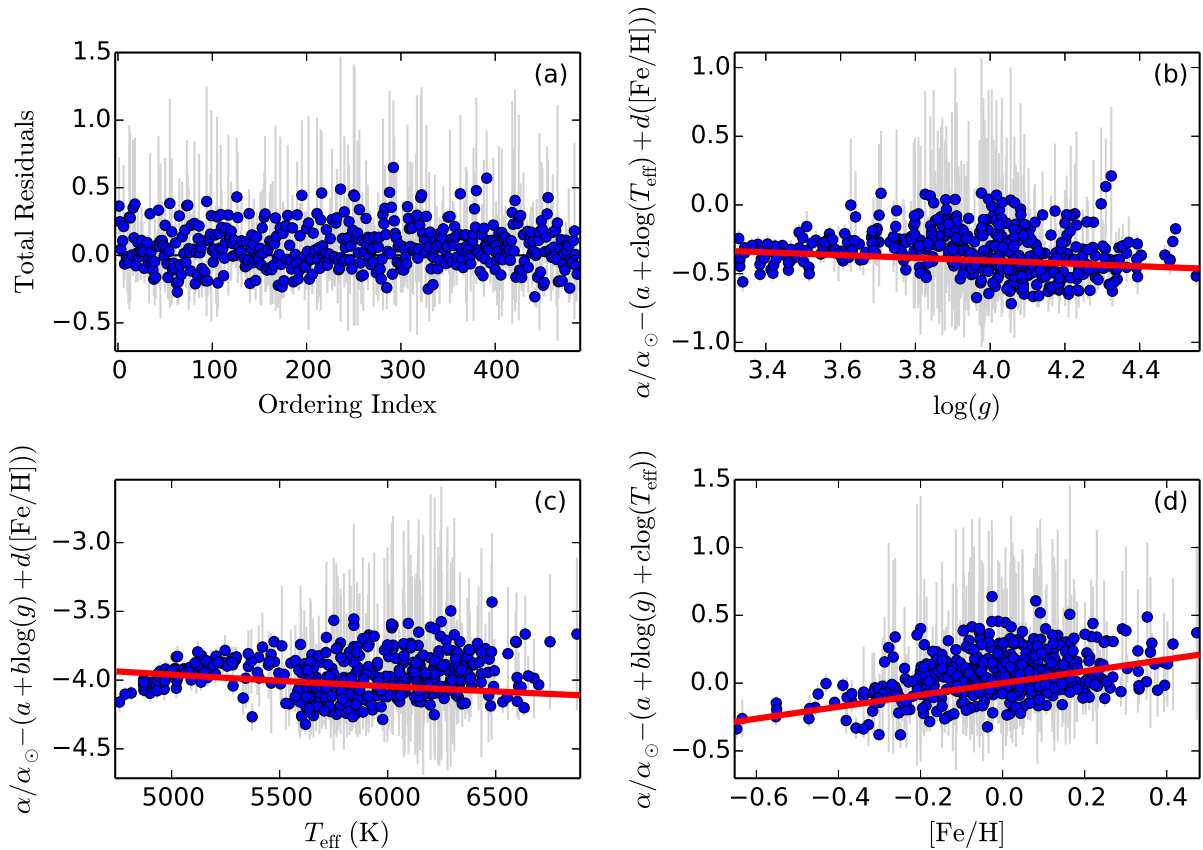}
\caption{The total residuals (a) and partial residuals (b-d) of the fit using Eq.~\ref{eq:trilinear_model} for the non-diffusion, $0.2H_p$ models. The red line in panel (b) it is $b \log(g)$, in (c) it is $c \log(T_{\mathrm{eff}})$, and in (d) it is $d \mathrm{[Fe/H]}$. The ordering index in panel (a) is simply the stars ordered by KIC number. 
}
\label{fig:Eddington_resids}
\end{figure*}

As in the analysis of \cite{Bonaca2012}, we perform a trilinear fit to model $\alpha / \alpha_{\sun}$ as a function of $\log(g)$, $\log(T_{\mathrm{eff}})$, and \FeH. Our equation takes the form,
\begin{equation}
\alpha/\alpha_{\sun}=a + b \log (g) + c \log (T_{\mathrm{eff}}) + d (\mathrm{[Fe/H]}).
\label{eq:trilinear_model}
\end{equation}
Using a minimum $\chi^2$ fit, the coefficients $a$, $b$, $c$, and $d$ for Eq.~\ref{eq:trilinear_model} were determined, with the results displayed in Table \ref{table:Eddington_coeffs}. Fig.~\ref{fig:Eddington_resids} shows the residuals and partial residuals for this fit. We find a positive trend between $\alpha/\alpha_{\sun}$ and \FeH, a negative trend between $\alpha/\alpha_{\sun}$ and $\log(T_{\mathrm{eff}})$, and a slightly negative trend between $\alpha/\alpha_{\sun}$ and $\log(g)$. 

\begin{table}
\centering
\caption{The best fit values to Eq.~\ref{eq:trilinear_model} for the non-diffusion, $0.2H_p$ models.}
\label{table:Eddington_coeffs}
\begin{tabularx}{9.87cm}{>{\raggedleft}p{2.1cm}>{\raggedleft}p{2.1cm}>{\raggedleft}p{2.1cm}>{\raggedleft\arraybackslash}p{2.1cm}}
\hline
\hline
\multicolumn{4}{c}{$\alpha/\alpha_{\sun}=a + b \log (g) + c \log (T_{\mathrm{eff}}) + d (\mathrm{[Fe/H]})$} \\ \hline
\multicolumn{1}{c}{$a$} & \multicolumn{1}{c}{$b$} & \multicolumn{1}{c}{$c$} & \multicolumn{1}{c}{$d$}  \\ \hline
5.426 $\pm$ 0.752  &   -0.101 $\pm$  0.025    & -1.071 $\pm$  0.221   & 0.437 $\pm$ 0.029 \\ \hline
\end{tabularx}
\end{table}

The best-fit coefficients to Eq.~\ref{eq:trilinear_model} for this work are compared to those found in \cite{Bonaca2012} in Table~\ref{table:compare_to_Bonaca}. Note that the coefficients presented in \cite{Bonaca2012} were for a fit to \Alpha\ and not $\alpha / \alpha_{\sun}$. To make the comparison with our results more clear, the \cite{Bonaca2012} values in Table~\ref{table:compare_to_Bonaca} have been divided by the value of $\alpha_{\sun}$ from \cite{Bonaca2012}. For each of the coefficients the sign is the same, however most notably the metallicity dependence from this work is larger. This is likely to be a result of the larger $\log g$ range and coverage of the current sample.

\begin{table}
\centering
\caption{The best fit values to Eq.~\ref{eq:trilinear_model} for the non-diffusion, $0.2H_p$ models and the results of \cite{Bonaca2012}.}
\label{table:compare_to_Bonaca}
\begin{tabularx}{14.973cm}{>{\raggedleft}p{2.2cm}|>{\raggedleft}p{2.2cm}>{\raggedleft}p{2.2cm}>{\raggedleft}p{2.2cm}>{\raggedleft\arraybackslash}p{2.2cm}}
\hline
\hline
\multicolumn{5}{c}{$\alpha/\alpha_{\sun}=a + b \log (g) + c \log (T_{\mathrm{eff}}) + d (\mathrm{[Fe/H]})$} \\ 
\hline
\multicolumn{1}{c}{Model Set} & \multicolumn{1}{c}{$a$} & \multicolumn{1}{c}{$b$} & \multicolumn{1}{c}{$c$} & \multicolumn{1}{c}{$d$}  \\ \hline
\multicolumn{1}{l}{Non-Diffusion, $0.2H_p$ Models} & 5.426 $\pm$ 0.752  &   -0.101 $\pm$  0.025    & -1.071 $\pm$  0.221   & 0.437 $\pm$ 0.029 \\
\multicolumn{1}{l}{Results of \cite{Bonaca2012}} & 4.72 $\pm$ 0.16$\phantom{1}$  &  -0.18 $\pm$ 0.05$\phantom{1}$  & -0.79 $\pm$ 0.47$\phantom{1}$  & 0.28 $\pm$ 0.07$\phantom{1}$ \\ \hline
\end{tabularx}
\end{table}

The properties of the stars in this study are shown in Table~\ref{table:star_data}. The values of $\nu_{\mathrm{max}}$ and $\Delta \nu$ were obtained from \cite{Serenelli2017}, the $T_{\mathrm{eff}}$ values are from \cite{MathurTemps}, the \FeH\ values are from \cite{Buchhave2015}, and the $\alpha$ values are the likelihood weighted average values for the non-diffusion, $0.2H_p$ models.

\begin{table}[]
\centering
\caption{The sample of stars in the study, a portion of the table is shown here, the full version is available online. The values of $\nu_{\mathrm{max}}$ and $\Delta \nu$ were obtained from \cite{Serenelli2017}, the $T_{\mathrm{eff}}$ values are from \cite{MathurTemps}, and the \FeH\ values are from \cite{Buchhave2015}. The corresponding $\alpha$ values are the likelihood weighted average values from this study for the non-diffusion, $0.2H_p$ models.}
\label{table:star_data}
\begin{tabular}{cccccc}
\hline
\hline
$\mathrm{KIC}$ & $\nu_{\mathrm{max}}$ ($\mu$Hz) & $\Delta \nu$ ($\mu$Hz)  & $T_{\mathrm{eff}}$ (K) & $\mathrm{[Fe/H]}$ & $\alpha$  \\ \hline
1430163  & 1775.247 $\pm$ $\phantom{1}$77.139  & 85.873 $\pm$ 1.882  & 6590  $\pm$ 50  & -0.05  $\pm$ 0.08 & 2.137 $\pm$ 0.614 \\ 
1435467 & 1382.311  $\pm$  $\phantom{1}$19.038 & 70.558   $\pm$ 0.087   & 6326  $\pm$ 50 & $\phantom{ }$ 0.01  $\pm$ 0.08 & 2.069 $\pm$ 0.237\\ 
1725815 & 1044.287  $\pm$ $\phantom{1}$54.759 & 55.942 $\pm$ 0.469  & 6330 $\pm$ 50   & -0.07  $\pm$ 0.08 & 1.670 $\pm$ 0.162 \\ 
2309595 & $\phantom{1}$643.208  $\pm$   $\phantom{1}$11.226 &  39.029  $\pm$ 0.721 & 5152 $\pm$ 50 & -0.09  $\pm$  0.08 & 1.944 $\pm$ 0.163 \\
2450729 & 1053.105 $\pm$ 114.904 & 61.910 $\pm$ 2.539 & 5868 $\pm$ 50  &  -0.24 $\pm$ 0.08 & 1.865
 $\pm$ 0.773 \\ 
\hline
\end{tabular}
\end{table}

\subsection{Trilinear Fit For Different Temperature Ranges and Evolutionary Phases}
\label{sec:Evolutionary_phase}
\begin{figure*}[h]
\epsscale{1.1}
\plotone{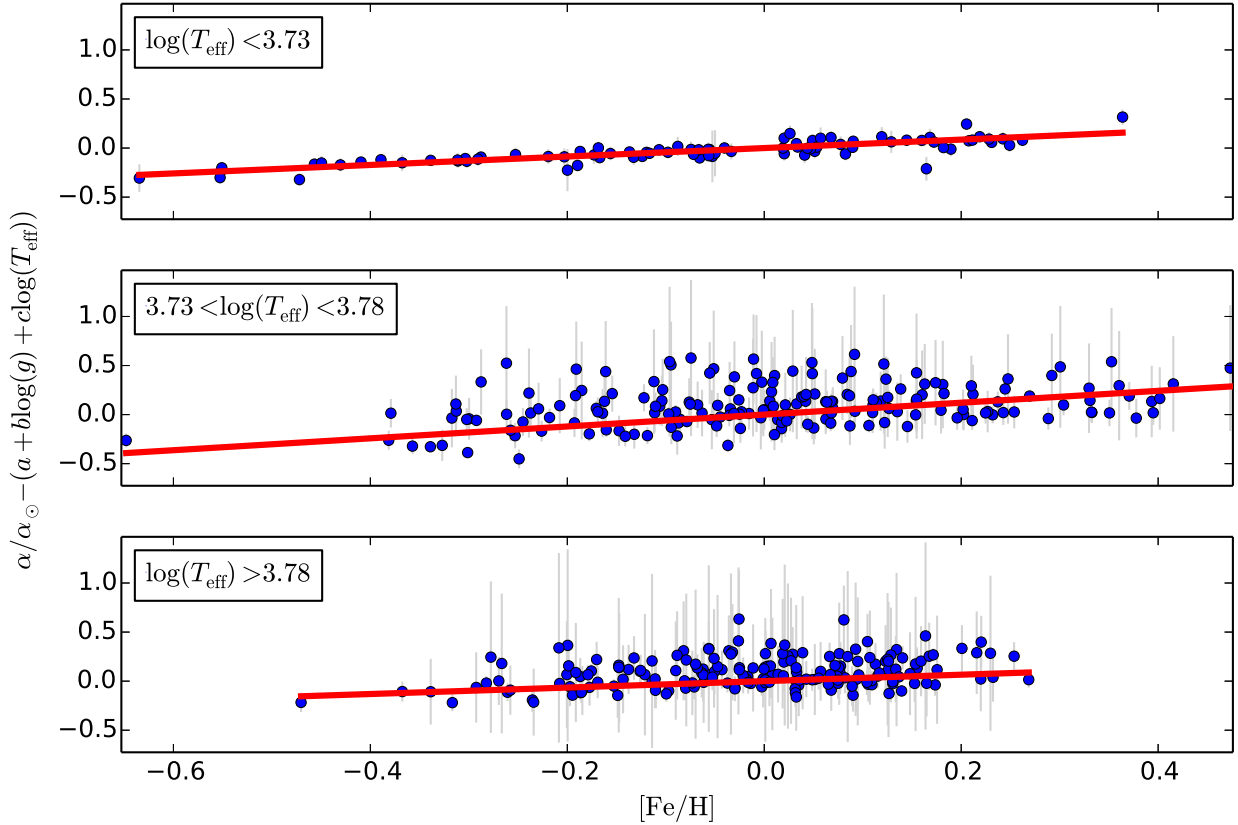}
\caption{The partial residuals as a function of \FeH\ of the fit using Eq.~\ref{eq:trilinear_model} for the non-diffusion, $0.2H_p$ models for each temperature range.
}
\label{fig:Temp_resids}
\end{figure*}

For the non-diffusion, $0.2H_p$ set of models, we also re-compute the trilinear fit (Eq.~\ref{eq:trilinear_model}) for 3 different temperature ranges, $\log(T_{\mathrm{eff}})<3.73$, $3.73<\log(T_{\mathrm{eff}})<3.78$, and $\log(T_{\mathrm{eff}})>3.78$. The partial residuals with respect to \FeH\ for these temperature divisions can be seen in Fig.~\ref{fig:Temp_resids} and the values of the fit coefficient can be found in Table~\ref{table:temp_divisions_coeff}. While the temperature range does affect the fit coefficients, we still see a positive correlation between \Alpha\ and \FeH\ in each range with the coefficient being between 0.328 and 0.605.

\begin{table}[h]
\centering
\caption{The best fit values and reduced $\chi^2$ values for the fit to Eq.~\ref{eq:trilinear_model} for various temperature range for the non-diffusion, $0.2H_p$ models.}
\label{table:temp_divisions_coeff}
\begin{tabularx}{16.5cm}{>{\raggedleft}p{2.2cm}>{\raggedleft}p{2.2cm}>{\raggedleft}p{2.2cm}>{\raggedleft}p{2.2cm}>{\raggedleft}p{2.2cm}>{\raggedleft\arraybackslash}p{1.5cm}}
\hline
\hline
\multicolumn{6}{c}{$\alpha/\alpha_{\sun}=a + b \log (g) + c \log (T_{\mathrm{eff}}) + d (\mathrm{[Fe/H]})$} \\ 
\hline
\multicolumn{1}{c}{Model Set}  & \multicolumn{1}{c}{$a$} & \multicolumn{1}{c}{$b$} & \multicolumn{1}{c}{$c$} & \multicolumn{1}{c}{$d$} & \multicolumn{1}{c}{$\chi^2$} \\ \hline
\multicolumn{1}{l}{Non-Diffusion, $0.2H_p$ (All $T_{\mathrm{eff}}$)} & 5.426 $\pm$ 0.752  & -0.101 $\pm$ 0.025 & -1.071 $\pm$  0.221  &  0.437 $\pm$ 0.029 & 1.333 $\phantom{a}$ \\ 
\multicolumn{1}{l}{$\log(T_{\mathrm{eff}})<3.73$} & -18.710 $\pm$ 2.339 & -0.183 $\pm$ 0.041  &   5.528 $\pm$ 0.652   &   0.431 $\pm$ 0.030 & 0.615 $\phantom{a}$ \\ 
\multicolumn{1}{l}{$3.73<\log(T_{\mathrm{eff}})<3.78$} & 22.639  $\pm$ 3.143 & 0.053 $\pm$  0.050  &  -5.830 $\pm$ 0.827 &  0.605  $\pm$ 0.052 & 1.156 $\phantom{a}$ \\ 
\multicolumn{1}{l}{$\log(T_{\mathrm{eff}})>3.78$} & -1.533 $\pm$ 2.695   &     -0.289 $\pm$ 0.047   &    0.969 $\pm$   0.702   &     0.328 $\pm$  0.059  & 0.693 $\phantom{a}$ \\ \hline
\end{tabularx}
\end{table}

We also test the impact of using a different temperature scale. The model likelihoods were recalculated, where $\mathcal{L}_{T_\mathrm{eff}}$ was determined using the ASPCAP $T_{\mathrm{eff}}$ values from \cite{Serenelli2017}. The trilinear fit was recomputed and the resulting $\mathrm{[Fe/H]}$ coefficient was 0.438 $\pm$ 0.031, compared to the original value of 0.437 $\pm$ 0.029 when using the original temperatures. So, the $\mathrm{[Fe/H]}$ coefficients are in excellent agreement. 

\begin{figure*}[h]
\epsscale{1.0}
\plotone{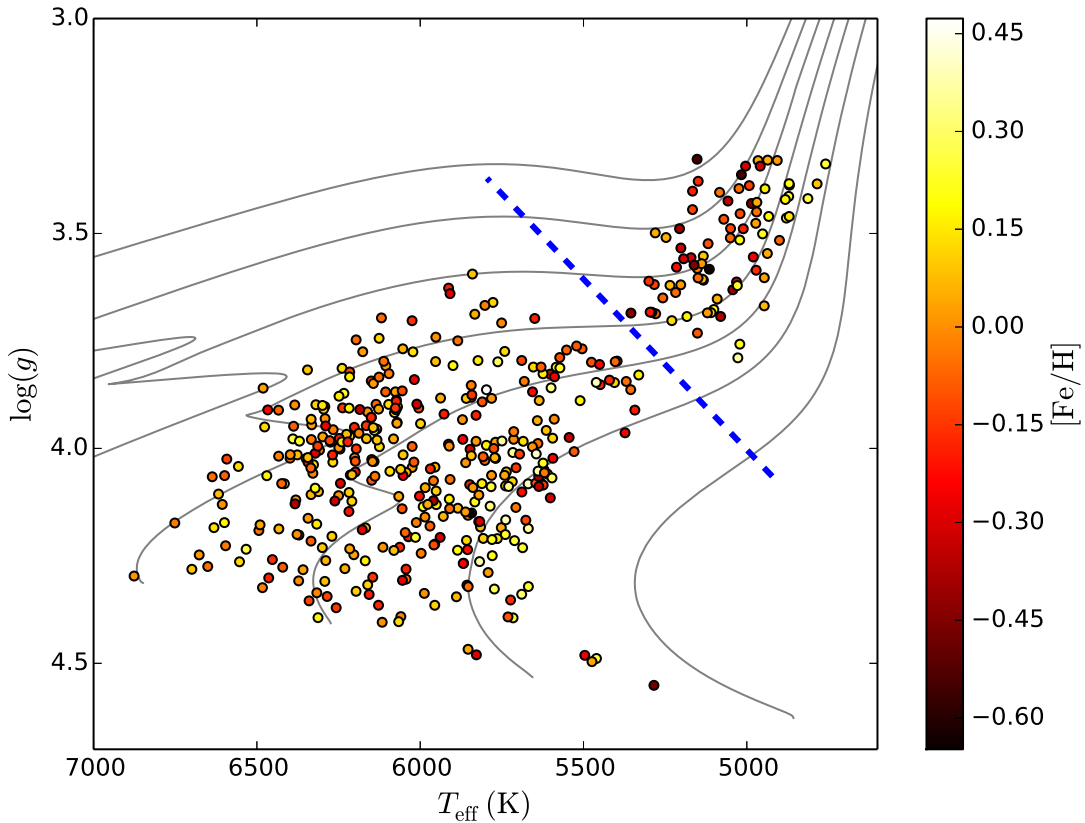}
\caption{A plot of $\log g$ vs. $T_{\mathrm{eff}}$ for the stars in the study. The points are colored by their likelihood weighted average value of \FeH. The dashed blue line separates the more evolved stars from the less evolved stars (used in Table~\ref{table:evolutionary_phase}). The background gray lines are tracks of 0.8, 1.0, 1.2, 1.4, 1.6, 1.8, and 2.0 $M_{\sun}$ created in YREC \citep{Demarque2008}, shown for reference.}
\label{fig:logg_T_plot}
\end{figure*}

We can also separate the stars by their evolutionary phase. Looking at the residuals from Fig.~\ref{fig:Eddington_resids}, the partial residuals for $T_{\mathrm{eff}}$ and $\log g$ appear to have different trends at low $T_{\mathrm{eff}}$ and low $\log g$, the more evolved stars, than compared to the partial residuals at higher $T_{\mathrm{eff}}$ and $\log g$. We can investigate if this is the result of a relation between the best fit coefficients, \Alpha, and evolutionary phase. Fig.~\ref{fig:logg_T_plot} plots the stars on the $\log g - T_{\mathrm{eff}}$ plane. As can be seen in Fig.~\ref{fig:logg_T_plot}, there is a distinct group of stars which are more evolved. This group of more evolved stars can also be seen in the H-R diagram in Fig.~\ref{fig:HR_plot}. We perform the trilinear analysis separately on the group of more evolved and less evolved stars to examine how the fit coefficients change, recorded in Table~\ref{table:evolutionary_phase}. Separating the stars into those which are more evolved and those which are less evolved does remove the features in the partial residuals at low $T_{\mathrm{eff}}$ and $\log g$. For the more evolved stars the \FeH\ coefficient decreased while in the less evolved stars the trend with \FeH\ increased. Additionally, for the more evolved stars, the $\log(T_{\mathrm{eff}})$ and $\log g$ coefficients changed sign.

\begin{table}
\centering
\caption{The best fit values and reduced $\chi^2$ values for the fit to Eq.~\ref{eq:trilinear_model} for the non-diffusion, $0.2H_p$ models, separated by evolutionary phase.}
\label{table:evolutionary_phase}
\begin{tabularx}{16.0cm}{>{\raggedleft}p{2.2cm}|>{\raggedleft}p{2.2cm}>{\raggedleft}p{2.2cm}>{\raggedleft}p{2.2cm}>{\raggedleft}p{2.2cm}>{\raggedleft\arraybackslash}p{1.0cm}}
\hline
\hline
\multicolumn{6}{c}{$\alpha/\alpha_{\sun}=a + b \log (g) + c \log (T_{\mathrm{eff}}) + d (\mathrm{[Fe/H]})$} \\ 
\hline
\multicolumn{1}{c}{Model Set} & \multicolumn{1}{c}{$a$} & \multicolumn{1}{c}{$b$} & \multicolumn{1}{c}{$c$} & \multicolumn{1}{c}{$d$} & \multicolumn{1}{c}{$\chi^2$}  \\ \hline
\multicolumn{1}{l}{Non-Diffusion, $0.2H_p$ Models} & 5.426 $\pm$ 0.752  &   -0.101 $\pm$  0.025    & -1.071 $\pm$  0.221   & 0.437 $\pm$ 0.029 & 1.333 \\
\multicolumn{1}{l}{More Evolved Stars} & -15.637  $\pm$ 2.086    &   0.022 $\pm$    0.052   &   4.504 $\pm$   0.591    &   0.390 $\pm$    0.026   & 0.448 \\
\multicolumn{1}{l}{Less Evolved Stars} & 4.523  $\pm$ 1.203    &  -0.050 $\pm$  0.037  &  -0.890 $\pm$   0.310   &   0.605 $\pm$  0.041  & 1.150 \\
\hline
\end{tabularx}
\end{table}

\subsection{Stars with Double Peaked Distributions in M and $\alpha$}
\begin{figure}[h]
\begin{center}
\includegraphics[width=2.5 in]{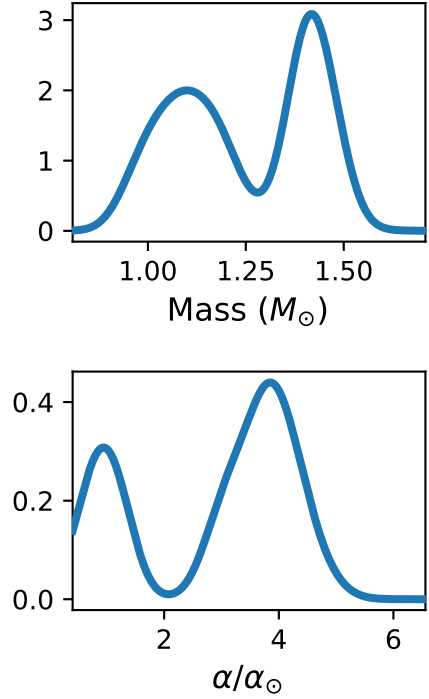}
\end{center}
\caption{An example of a star, KIC 2010607, with a bimodal PDF in both $M$ and \Alpha.}
\label{fig:bimodal_example}
\end{figure}

\begin{figure}[h]
\begin{center}
\includegraphics[width=2.5in]{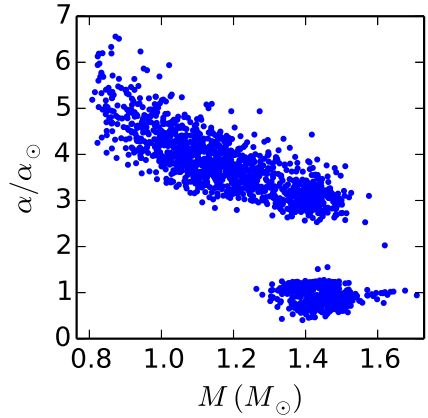}
\end{center}
\caption{A plot of $\alpha/\alpha_{\sun}$ as a function of mass for an example star (KIC 2010607) which had a bimodal PDF in both $M$ and \Alpha. Compared to the PDF of this star as seen in Fig.~\ref{fig:bimodal_example}, it can be seen that the lower mass peak corresponds to the higher $\alpha / \alpha_{\sun}$ peak and vice versa.}
\label{fig:multi_example_m_vs_alpha}
\end{figure}

An examination of the PDF and model likelihood results for each star shows that some stars have bimodal distributions both in mass and \Alpha. An example of such a PDF is that of KIC 2010607, which is shown in Fig.~\ref{fig:bimodal_example}. In these cases, the lower mass peak corresponds to the higher \Alpha\ peak and vice versa. This can be seen in Fig.~\ref{fig:multi_example_m_vs_alpha}. About 5\% of the stars have such a bimodal distribution in both mass and \Alpha.  

In these cases, a bimodal Gaussian function was fit to the PDF histogram to determine the value of each peak and the corresponding $\sigma$. The star was then treated as having two separate solutions, one for each set of mass and \Alpha\ peaks. We can examine how separating the peaks effects the results compared to the original values for the non-diffusion, $0.2H_p$ models. With the bimodal stars split the \FeH\ coefficient is 0.528 $\pm$ 0.034 and with the bimodal stars removed from the sample the \FeH\ coefficient is 0.436 $\pm$ 0.030, compared to the original \FeH\ coefficient of 0.437 $\pm$ 0.029.

\subsection{Analysis with Alternate Model Physics}
\label{sec:other_physics}

\begin{figure*}[h]
\epsscale{1.1}
\plotone{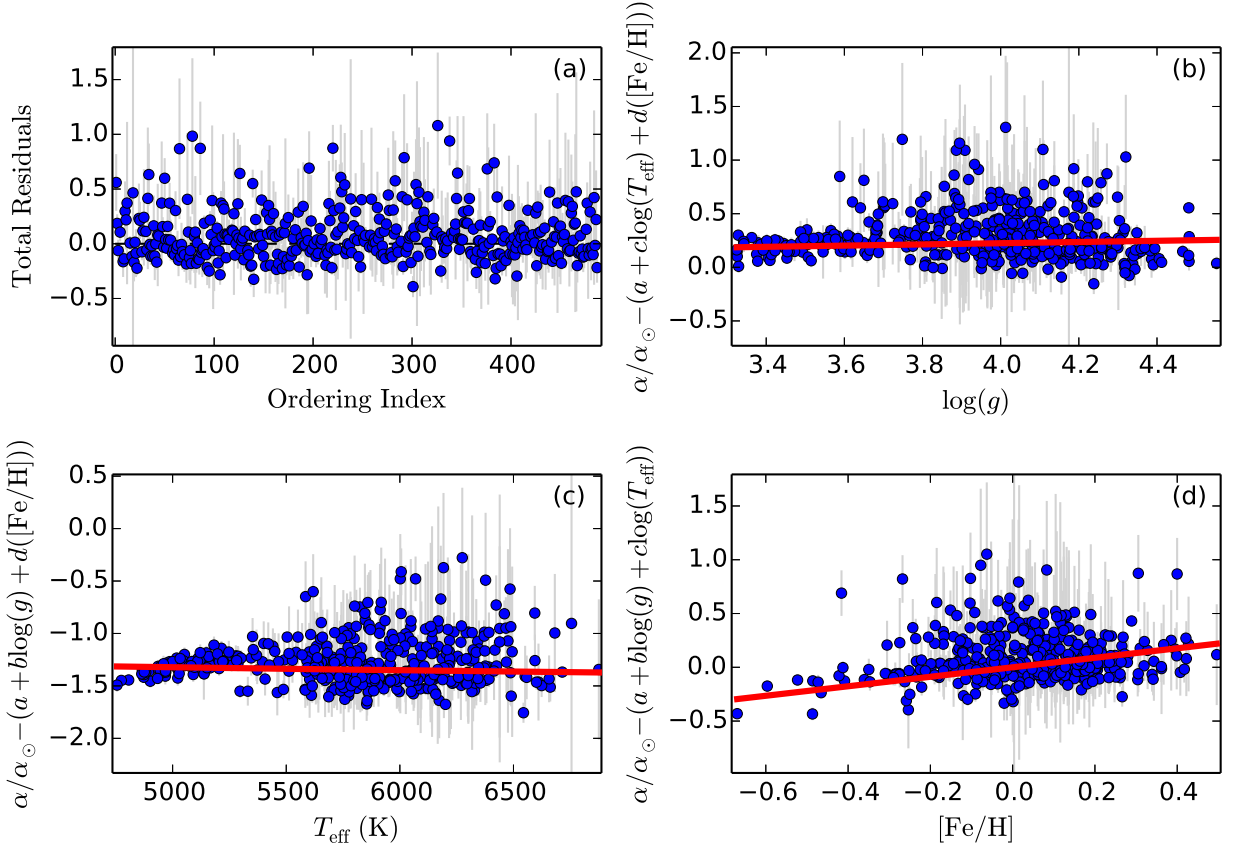}
\caption{The total residuals (a) and partial residuals (b-d) of the fit using Eq.~\ref{eq:trilinear_model} for the models with diffusion. The red line in panel (b) it is $b \log(g)$, in (c) it is $c \log(T_{\mathrm{eff}})$, and in (d) it is $d \mathrm{[Fe/H]}$.
}
\label{fig:diffusion_resids}
\end{figure*}

\begin{figure*}[h]
\epsscale{1.1}
\plotone{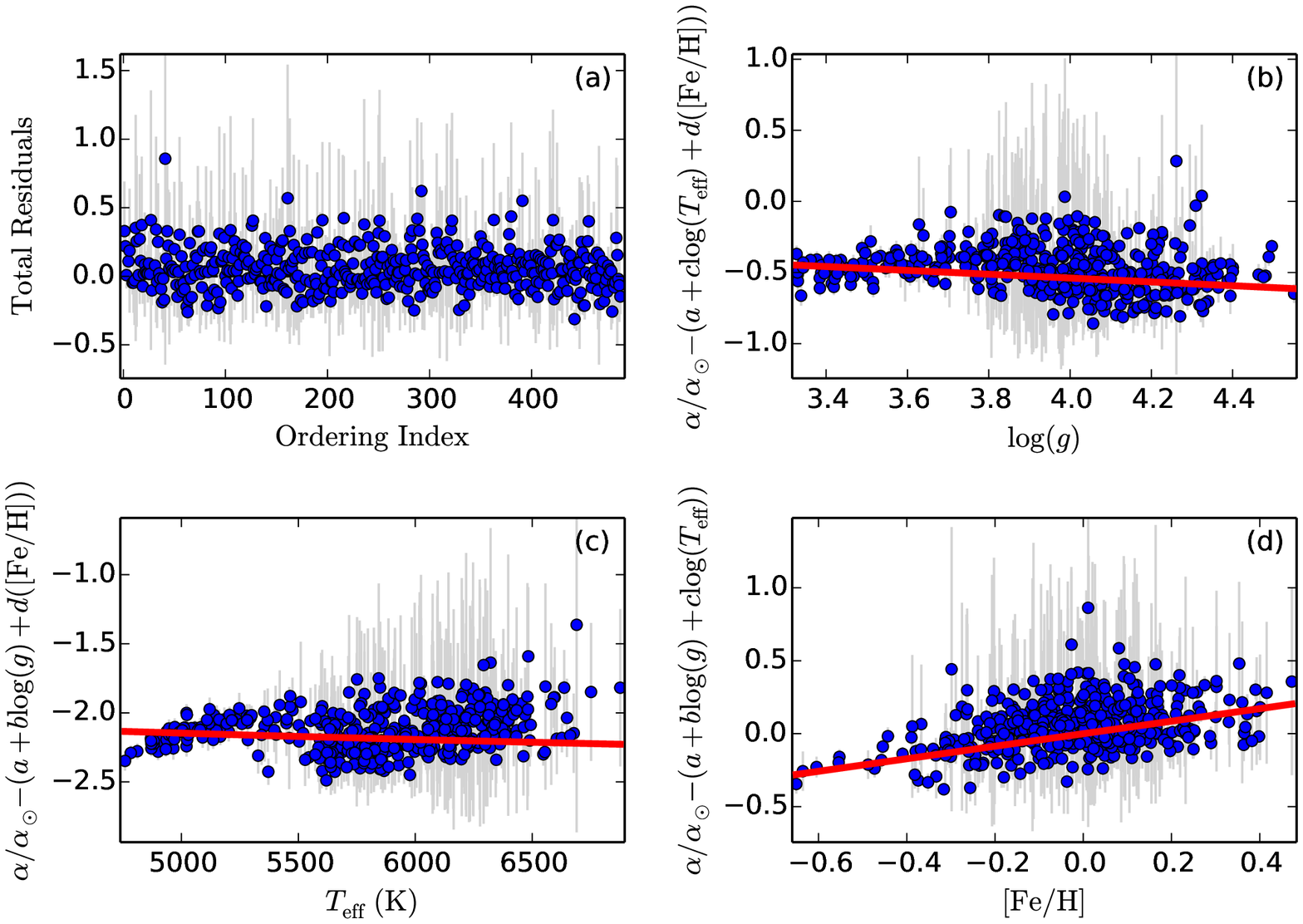}
\caption{The total residuals (a) and partial residuals (b-d) of the fit using Eq.~\ref{eq:trilinear_model} for the models without the inclusion of the $\mathcal{L}_{\nu_{\mathrm{max}}}$ term in the likelihood calculation. The red line in panel (b) it is $b \log(g)$, in (c) it is $c \log(T_{\mathrm{eff}})$, and in (d) it is $d \mathrm{[Fe/H]}$.
}
\label{fig:no_numax_resids}
\end{figure*}

The procedure was repeated with several other sets of stellar models, all with unique model physics. The additional sets were models including diffusion, a set of models without diffusion and without overshoot, and a recalculation of the original non-diffusion, $0.2H_p$ models without the inclusion of the $\mathcal{L_{\nu_{\mathrm{max}}}}$ term in the likelihood calculation. For each set of stellar physics, the best fit values for the coefficients in the trilinear fit of Eq.~\ref{eq:trilinear_model} are shown in Table~\ref{table:trilinear_results_all}. The residuals and partial residuals for the other sets of models can be seen in Figs.~\ref{fig:diffusion_resids} and \ref{fig:no_numax_resids}. In all cases, the linear model agrees well with the data. It is interesting to note that the \FeH\ coefficients for the different model sets are in agreement with each other while the $\log g$ and $\log(T_{\mathrm{eff}})$ coefficients are not. The $\log(T_{\mathrm{eff}})$ coefficient changes in magnitude fairly substantially and the $\log g$ coefficient changes in sign depending on the model physics while the correlation with \FeH\ appears to be more model independent.

\begin{table}
\centering
\caption{The best fit values and reduced $\chi^2$ for the fit to Eq.~\ref{eq:trilinear_model} for the different sets of stellar physics.}
\label{table:trilinear_results_all}
\begin{tabularx}{16.25cm}{>{\raggedleft}p{2.2cm}>{\raggedleft}p{2.2cm}>{\raggedleft}p{2.2cm}>{\raggedleft}p{2.2cm}>{\raggedleft}p{2.2cm}>{\raggedleft\arraybackslash}p{1.5cm}}
\hline
\hline
\multicolumn{6}{c}{$\alpha/\alpha_{\sun}=a + b \log (g) + c \log (T_{\mathrm{eff}}) + d (\mathrm{[Fe/H]})$} \\ 
\hline
\multicolumn{1}{c}{Model Set} & \multicolumn{1}{c}{$a$} & \multicolumn{1}{c}{$b$} & \multicolumn{1}{c}{$c$} & \multicolumn{1}{c}{$d$} & \multicolumn{1}{c}{$\chi^2$} \\ \hline
\multicolumn{1}{l}{Non-Diffusion, $0.2H_p$ Models} & 5.426 $\pm$ 0.752  &   -0.101 $\pm$  0.025    & -1.071 $\pm$  0.221   & 0.437 $\pm$ 0.029 & 1.333 $\phantom{a}$ \\  
\multicolumn{1}{l}{Diffusion Models} & 2.162 $\pm$  0.463  &   0.056 $\pm$  0.017  &   -0.357 $\pm$   0.134   &    0.441 $\pm$  0.027 & 1.948 $\phantom{a}$ \\  
\multicolumn{1}{l}{No $\nu_{\mathrm{max}}$} & 3.728 $\pm$ 0.783  & -0.135 $\pm$ 0.025  & -0.580 $\pm$ 0.229  & 0.429 $\pm$ 0.027 & 1.160 $\phantom{a}$ \\ 
\multicolumn{1}{l}{Non-Diffusion, No Overshoot} & 9.546 $\pm$  1.293 & 0.024  $\pm$  0.035  &  -2.306 $\pm$ 0.376 & 0.410 $\pm$ 0.036 &  1.615 $\phantom{a}$ \\ \hline
\end{tabularx}
\end{table}

$ $

$ $

\subsubsection{Comparing Results with and without Diffusion} 
\label{sec:diff_compare}
\begin{figure*}[h]
\epsscale{1.1}
\plotone{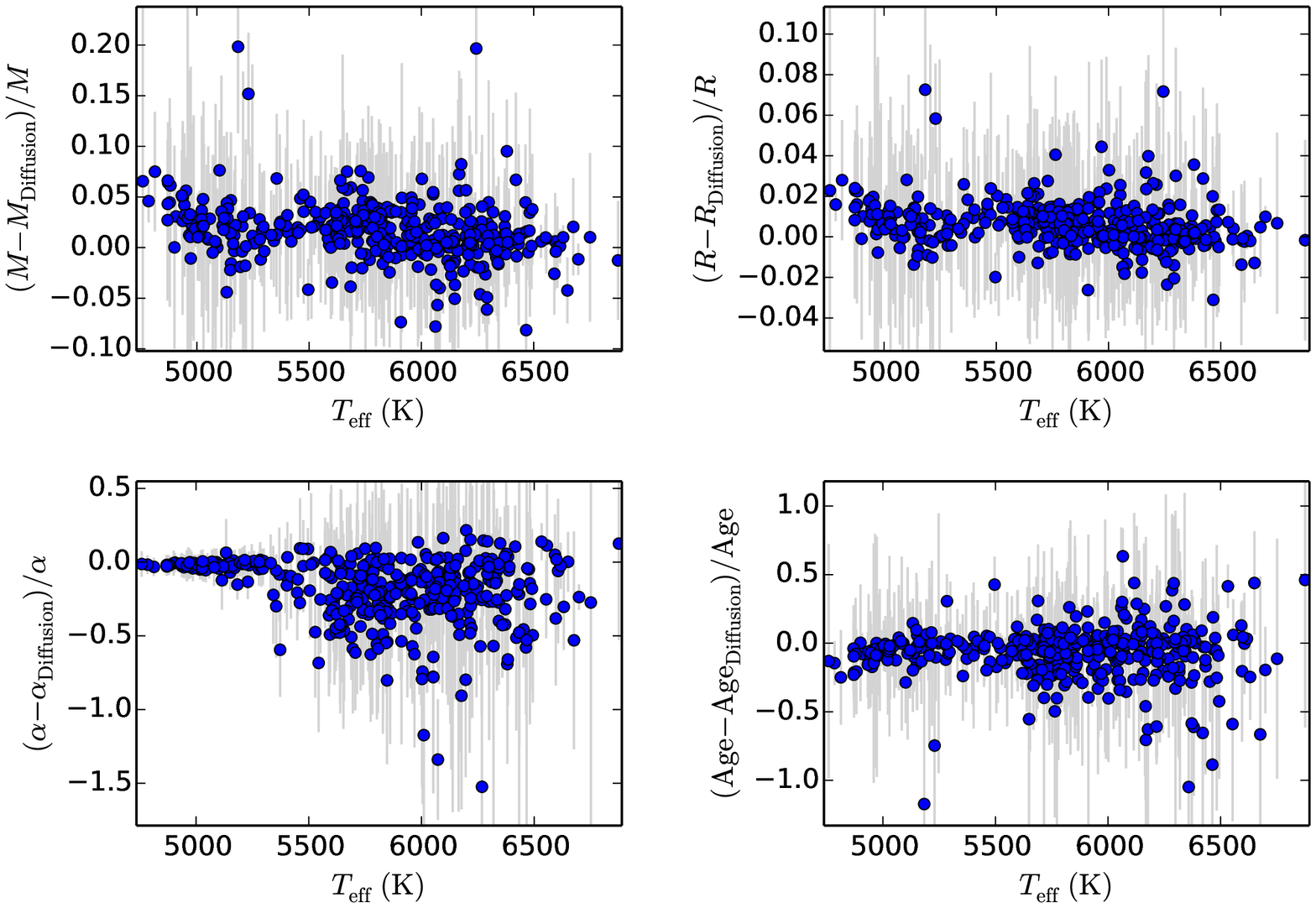}
\caption{Comparing each star's likelihood weighted average value for $M$, $R$, $\alpha$, and age for the models created with and without diffusion.}
\label{fig:diffusion_compare}
\end{figure*}

We can examine the effect that including diffusion had on the likelihood weighted average values for each star. Figure~\ref{fig:diffusion_compare} compares the fractional difference in $M$, $R$, \Alpha, and age for each star both with and without diffusion. As can be seen in Fig.~\ref{fig:diffusion_compare} the fractional mass difference is less than 10\% for all but a few stars and the fractional radius difference is less than 4\% for all but a few stars. The typical fractional difference in $T_{\mathrm{eff}}$ is less than 0.5\%, the fractional differences in $\Delta \nu$ and $\nu_{\mathrm{max}}$ is around or less that 5\%, and the difference in \FeH\ is less than 0.1 dex in almost all cases. 

\subsubsection{Comparing Results with and without the $\mathcal{L_{\nu_{\mathrm{max}}}}$ Term} 
\label{sec:no_numax}
For the non-diffusion, $0.2H_p$ set of models, the likelihoods were calculated both with and without the $\mathcal{L_{\nu_{\mathrm{max}}}}$ term in Eq.~\ref{eq:likelihoods_total}. The underlying theory explaining the physical mechanism which gives rise to $\nu_{\mathrm{max}}$ is not completely understood. While the relation between $\nu_{\mathrm{max}}$ and the properties of a star can be approximated through the scaling relation and while $\nu_{\mathrm{max}}$ can be shown to be proportional to $\nu_{\mathrm{ac}}$, the reasons which give rise to these relationships are still not fully explained. As a result of this incomplete understanding of $\nu_{\mathrm{max}}$, here we recalculate the results of this work if the $\nu_{\mathrm{max}}$ term in the likelihood calculation is ignored. We examine the effect that the omission of the $\mathcal{L_{\nu_{\mathrm{max}}}}$ term had on the likelihood weighted average values of each star. It is important to note here that although the $\mathcal{L_{\nu_{\mathrm{max}}}}$ term was not used in calculating the likelihoods, $\nu_{\mathrm{max}}$ was a prior in the Monte Carlo and so $\nu_{\mathrm{max}}$ information is not truly being completely ignored.

The values of the likelihood weighted average stellar properties do not change greatly if the $\mathcal{L_{\nu_{\mathrm{max}}}}$ term is included or not. The fractional difference in the likelihood weighted average for $T_{\mathrm{eff}}$ is less than 0.5\%, for $\Delta \nu$ and $\nu_{\mathrm{max}}$ the fractional difference is around or less than 5\%, the difference in \FeH\ is less than 0.05 dex for almost all cases, the fractional mass difference is less than 2\% for most stars and less than 10\% even in the most extreme cases, the fractional radius difference is less than 2\%, less than 3\% difference in \Alpha\ for the vast majority of the stars, and less than 5\% age difference for most stars. So, the effects of omitting the $\nu_{\mathrm{max}}$ term from the likelihood calculations is minor. This can also be seen in Table~\ref{table:trilinear_results_all} as the metallicity coefficient changed by less than 2\% with the omission of the $\mathcal{L_{\nu_{\mathrm{max}}}}$ term.

\subsubsection{Comparing the Results with and without Overshoot} 
Similarly, we can compare the differences in the likelihood weighted averages for the non-diffusion models with and without overshoot. The fractional difference in the likelihood weighted average for \teff\ are less than 0.5\%, for \dnu\ and \numax\ the fractional difference is about 1\%. The difference between the \FeH\ values is less than 0.01 dex for the majority of the stars. The fractional difference in radius is 2\% and for mass it is about 4\%. The fractional difference for \Alpha\ is less than 5\% for the majority of the stars. For the vast majority of the points the fractional difference in age is about 10\%.

\section{Discussion and Conclusions}
\label{sec:Discussion}

\subsection{Effect on Isochrones}
If the value of \Alpha\ does depend on the metallicity of the star, then for a given metallicity the temperature-luminosity relation changes. This in turn will change the isochrones, especially on the giant branch, as mentioned in \cite{Demarque1992}, for example. \cite{Yi2003} explains that since a larger \Alpha\ makes convection more efficient, then the stellar model will be bluer and hotter. This color uncertainty can cause age uncertainties of 25\% \citep{Yi2003}.

To demonstrate the effects that a metallicity dependent \Alpha\ has on stellar isochrones, isochrones were created for metallicities of $\mathrm{[Fe/H]}=-0.5$ and $+0.5$ for the non-diffusion, $0.2H_p$ set of models. For each metallicity two sets of isochrones were created in YREC, one set with the solar calibrated value of \Alpha, and one set with the value of \Alpha\ following the trend observed between \Alpha\ and \FeH. The $\mathrm{[Fe/H]}=-0.5$ isochrones are 8 Gyr while the $\mathrm{[Fe/H]}=+0.5$ isochrones are 1 Gyr in age.

\begin{figure}[h]
\begin{center}
\includegraphics[width=3.0 in]{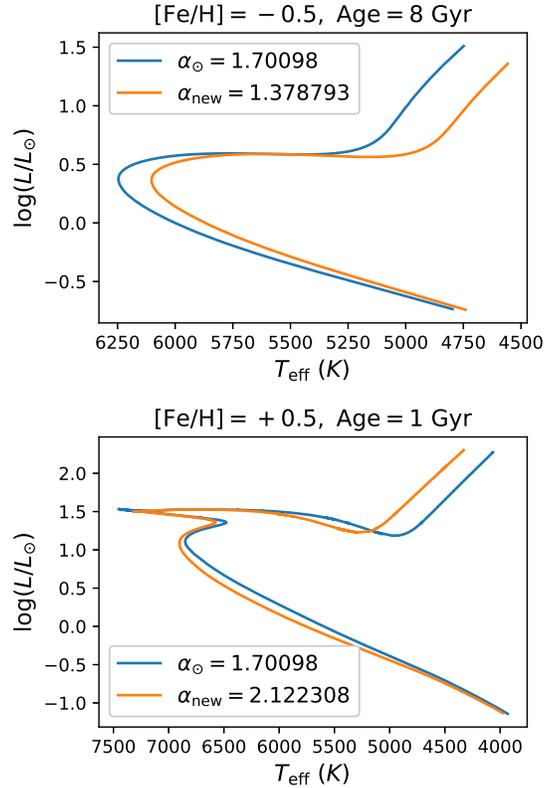}
\end{center}
\caption{Isochrones showing the differences between using the solar calibrated value of \Alpha\ (blue lines) and the metallicity dependent value of \Alpha\ (orange lines) determined from this work. The top panel shows 8 Gyr isochrones for $\mathrm{[Fe/H]}=-0.5$ while the bottom panel shows 1 Gyr isochrones for $\mathrm{[Fe/H]}=+0.5$.}
\label{fig:isochrones}
\end{figure}

These isochrones can be seen in Fig.~\ref{fig:isochrones}. As can be seen in Fig.~\ref{fig:isochrones}, a smaller mixing length parameter shifts the isochrones towards cooler temperatures, agreeing with the explanation given by \cite{Yi2003}. The clear difference between the sets of isochrones with different \Alpha\ values, especially at turn-off and the giant branch, show the importance of correctly selecting the value of the mixing length parameter as opposed to relying on the the solar calibrated value.

\subsection{Testing Other Functional Forms}
\begin{figure*}[h]
\epsscale{1.1}
\plotone{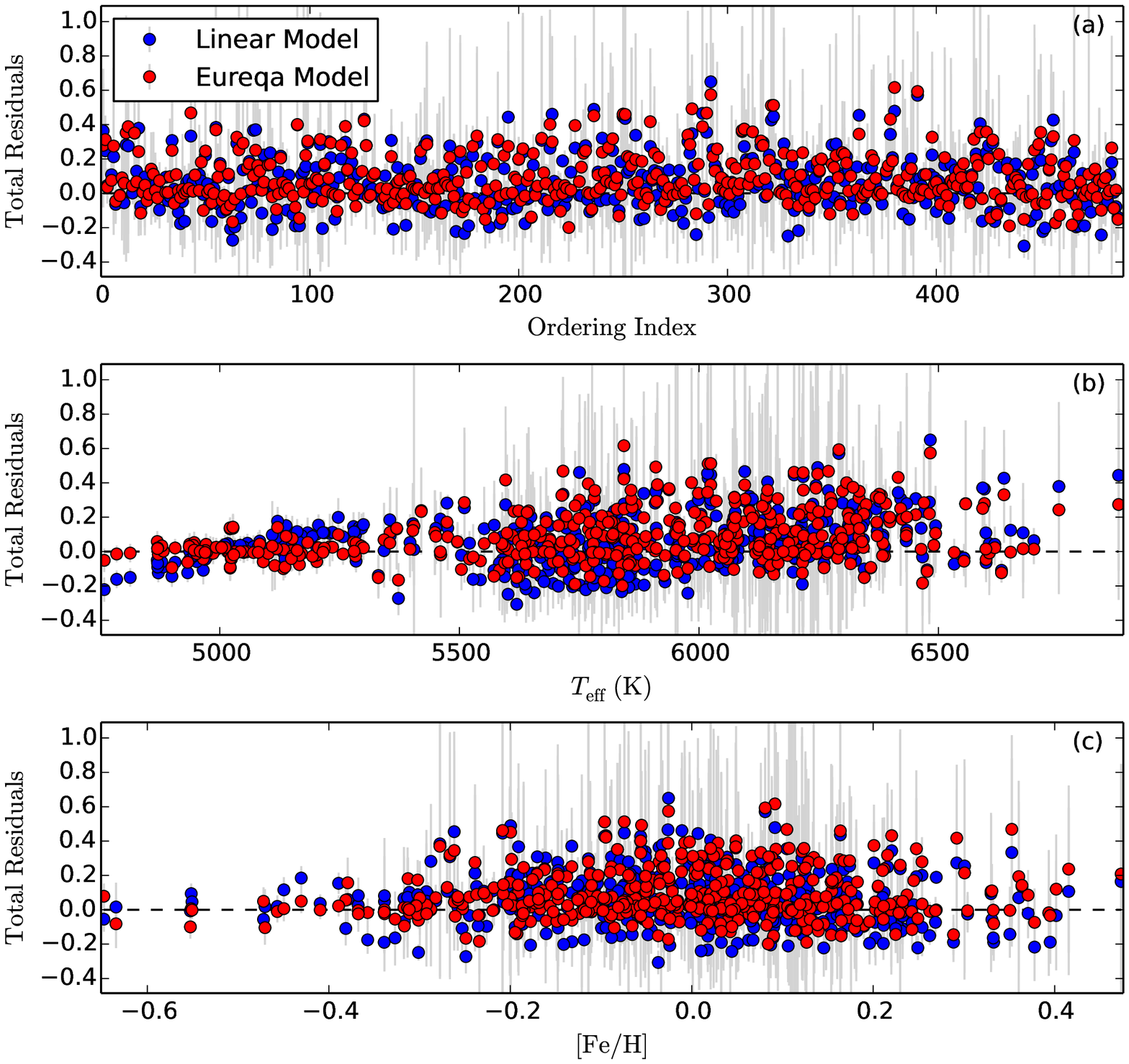}
\caption{The residuals for the linear model from Eq.~\ref{eq:trilinear_model} (blue points) compared to the more complex Eq.~\ref{eq:Eureqa_fit} determined using the \textit{Eureqa} software (red points). The residuals are shown as a function of the ordering index (Panel a), as a function of $T_{\mathrm{eff}}$ (Panel b), and as a function of \FeH\ (Panel c).}
\label{fig:Eureqa_compare}
\end{figure*}

It is also possible that the relationship between \Alpha, \teff, \logg, and \FeH\ is not a linear one, but takes on some other functional form. To investigate this, the software package \textit{Eureqa} \citep{SchmidtLipson2009} was used. \textit{Eureqa}, available from the Nutonian company, performs symbolic regression through the use of an evolutionary search. The evolutionary search found that for the non-diffusion, $0.2H_p$ models the best fit took the form,
\begin{equation}
\begin{split}
\frac{\alpha}{\alpha_{\sun}} = 1.0477 + \frac{0.0002}{\mathrm{[Fe/H]} - 0.1119} + 0.0103 \mathrm{[Fe/H]} \cos(-13.5879 \log(T_{\mathrm{eff}})) \exp(\log(g)) \\
 - 0.2339 \sin(4.9127 \log(g)) \sin(-21.3015 \log(T_{\mathrm{eff}})).
\end{split}
\label{eq:Eureqa_fit}
\end{equation}
We can compare the linear fit from Eq.~\ref{eq:trilinear_model} to this more complicated equation, as seen in Fig.~{\ref{fig:Eureqa_compare}}, which shows the total residuals, $(\alpha / \alpha_{\sun})_{\mathrm{Model \: Stars}} - (\alpha / \alpha_{\sun})_{\mathrm{Equational \: Fits}}$.
From Fig.~{\ref{fig:Eureqa_compare}, the \textit{Eureqa} functional form and the linear model residuals do not appear significantly different. It is worth noting that in the low temperature region the \textit{Eureqa} model residuals are much closer to zero compared to the residuals for the linear model. Comparing the reduced $\chi^2$ values for the linear model and the \textit{Eureqa} equation we find that $\chi_{\mathrm{Linear}}^2 = 1.333$ and $\chi_{\mathrm{Eureqa}}^2 = 0.615$. While the $\chi^2$ value is about a factor of 2 better, the complicated form of Eq.~\ref{eq:Eureqa_fit} does not appear to be justified on any theoretical basis. While the trilinear fit is not either, it is at least a simple equational form which provides a good fit.

The inclusion of a $\log(M)$ term in Eq.~\ref{eq:trilinear_model} was also tested. A quadrilinear fit including $\log(T_{\mathrm{eff}})$, $\log(g)$, \FeH, and $\log(M)$ was performed for the non-diffusion, $0.2H_p$ models. The resulting fit coefficient for the \FeH\ term was 0.463 $\pm$ 0.039 for the quadrilinear fit, compared to 0.437 $\pm$ 0.029 for the trilinear fit without the $\log(M)$ term. The reduced $\chi^2$ for the quadrilinear fit was 1.335, compared to 1.333 for the trilinear fit.

The analysis for the non-diffusion, $0.2H_p$ models was also repeated using principal component analysis (PCA). PCA was used to transform the variables \logg, \FeH, and $\log(T_{\mathrm{eff}})$ into three orthogonal principal components in the directions of the the greatest variance. These three principal components were then fit linearly to $\alpha / \alpha_{\odot}$. The resulting fit, when converted back to the original variables instead of the principal components, gives the same fit coefficients as previously determined using the original linear fit.

\subsection{Comparison with Other Work}

Our results can be compared to other works in literature. \cite{Tayar2017} compared stellar models and stars in the APOKASC sample \citep{Pinsonneault2014}, finding a metallicity dependent temperature offset. \cite{Tayar2017} demonstrated that a metallicity dependent mixing length, of the form,  $\alpha = 0.1612 \mathrm{[Fe/H]} + 1.9037$ improves the differences. For the sake of comparison we divide the \cite{Tayar2017} correction by their value of $\alpha_{\sun}$ so that $\alpha / \alpha_{\sun} = 0.0937 \mathrm{[Fe/H]} + 1.1068$. For our sample, the line of best fit for $\alpha / \alpha_{\sun}$ as a function of \FeH\ gives: $\alpha / \alpha_{\sun} = 0.437 \mathrm{[Fe/H]} + 1.029$. So, we find a much stronger metallicity dependence in this set of stars. However, \cite{Tayar2017} performed their analysis on red giant stars, while our sample consists of mainly dwarfs  and subgiants. As demonstrated in Sec.~\ref{sec:Evolutionary_phase}, the trend between \Alpha\ and \FeH\ is weaker in more evolved stars. So, we would expect the \FeH\ coefficient for the \cite{Tayar2017} red giant stars to be smaller compared to this sample. Taking only the more evolved stars in this work (see Fig.~\ref{fig:logg_T_plot}) then the line of best fit is: $\alpha / \alpha_{\sun} =  0.292 \mathrm{[Fe/H]} + 1.109$.

\cite{Metcalfe2014} modeled 42 \textit{Kepler} target stars using the Asteroseismic Modeling Portal \citep{Metcalfe2009,Woitaszek2009} and found a negative correlation between \Alpha\ and $T_{\mathrm{eff}}$ and a positive trend between \Alpha\ and \FeH\ and $g$. Additinally, \cite{Creevey2017} modeled 57 stars using the Asteroseismic Modeling Portal and also found a positive trend between \Alpha\ and metallicity as well as between \Alpha\ and $\log g$ and a negative trend between \Alpha\ and $\log (T_{\mathrm{eff}})$. This trend between \Alpha\ and $T_{\mathrm{eff}}$ and \FeH\ agrees with our results, however the trend with surface gravity does not. This, however, could be a result of the larger range of $\log g$ in our sample. If our range of $\log g$ is restricted to the same range as \cite{Creevey2017} then we also see a positive correlation between \Alpha\ and $\log g$ and \Alpha\ and metallicity with a negative trend between \Alpha\ and \teff.

It is much more interesting to compare these results with those obtained from 3D simulations. However, the complicating issue there is that defining \Alpha\ for a 3D simulation is difficult, and the definitions are ambiguous given that 3D simulations show that convection looks nothing like the MLT picture of overturning eddies of a given size. Nonetheless, there have been attempts to define \Alpha\ from the simulations. \cite{Magic2015} did a grid of 3D radiative hydrodynamic simulations of convection at different \logg, \teff, and metallicity and
produced three different types of \Alpha\ that we could fit to Eq.~\ref{eq:trilinear_model} and compare to our results. The three \Alpha\ values that \cite{Magic2015} produced  are $\alpha_{\mathrm{MLT}}^{s_\mathrm{bot}}$, $\alpha_{\mathrm{MLT}}^{\Delta s}$, and $\alpha_{\mathrm{m}}$. As explained in \cite{Magic2015}, the two different forms of $\alpha_{\mathrm{MLT}}$ are computed and calibrated by matching $s_{\mathrm{bot}}$ or $\Delta s$ between the 1D and 3D models. Here $\Delta s$ is the entropy jump, defined as $\Delta s = s_{\mathrm{bot}}-s_{\mathrm{min}}$ and $s_{\mathrm{bot}}$ is the asymptotic entropy of the deep convective region. The third \Alpha\ value, $\alpha_{\mathrm{m}}$, is the mass mixing length, which \cite{Magic2015} define as the inverse gradient of the vertical mass flux. We then perform the trilinear fit to our Eq.~\ref{eq:trilinear_model} and compare the coefficients for these different \Alpha\ values to our results. The best-fit coefficient for each different \Alpha\ value can be seen in Table~\ref{table:simulation_compare} where \cite{Magic2015} is referred to as \citetalias{Magic2015}. Mixing length values of models were also obtained from \cite{Trampedach2014} (henceforth \citetalias{Trampedach2014}), and the trilinear fit was also performed on these stars as well as on values of $\alpha_{\mathrm{m}}$ obtained from \cite{Trampedach2011} (henceforth \citetalias{Trampedach2011}). The
\cite{Trampedach2011} and \cite{Trampedach2014} models were all of solar metallicity, and hence the \FeH\ coefficient in Eq.~\ref{eq:trilinear_model} is not determined for these cases.

As with the \cite{Tayar2017} comparison with simulations, the metallicity-dependence of our mixing length results show a disagreement when compared to ``mixing lengths'' obtained from  3D simulations of convection; the sign of the change is the opposite. The reasons are not completely clear. The real $T$--$\tau$ relation in stars is more complicated than the Eddington $T$--$\tau$ relation, but \cite{Tanner2014} showed that an $\alpha$ that increases with metallicity is also obtained when models are constructed with the $T$--$\tau$ relation obtained from convection simulations of the correct metallicity.

\begin{table}[h]
\centering
\caption{The best-fit values to Eq.~\ref{eq:trilinear_model} for the non-diffusion, $0.2H_p$ models and results from various convection simulation studies.}
\label{table:simulation_compare}
\begin{tabularx}{15.29cm}{>{\raggedleft}p{2.1cm}>{\raggedleft}p{2.1cm}>{\raggedleft}p{2.3cm}>{\raggedleft}p{2.1cm}>{\raggedleft}p{2.1cm}>{\raggedleft\arraybackslash}p{2.1cm}}
\hline
\hline
\multicolumn{6}{c}{$a + b \log (g) + c \log (T_{\mathrm{eff}}) + d (\mathrm{[Fe/H]})$} \\ 
\hline
\multicolumn{1}{c}{Fitted Parameter} & \multicolumn{1}{r}{Source} & \multicolumn{1}{c}{$a$} & \multicolumn{1}{c}{$b$} & \multicolumn{1}{c}{$c$} & \multicolumn{1}{c}{$d$}  \\ \hline
\multicolumn{1}{l}{$\alpha / \alpha_{\sun}$ } & This Work & 5.426 $\pm$ 0.752  &   -0.101 $\pm$  0.025    & -1.071 $\pm$  0.221   & 0.437 $\pm$ 0.029  \\
\multicolumn{1}{l}{$\alpha_{\mathrm{m}}/\alpha_{\mathrm{m},\sun}$} & \citetalias{Magic2015} & 12.810 $\pm$ 1.292  & 0.177 $\pm$ 0.022  & -3.355 $\pm$ 0.359   & -0.072    $\pm$ 0.013   \\
\multicolumn{1}{l}{$\alpha_{\mathrm{MLT}}^{S_{\mathrm{bot}}} / \alpha_{\mathrm{MLT,\sun}}^{S_{\mathrm{bot}}}$ } & \citetalias{Magic2015} & 4.065 $\pm$ 0.201 & 0.058 $\pm$ 0.004  & -0.885 $\pm$ 0.056 & -0.004 $\pm$ 0.002 \\              
\multicolumn{1}{l}{$\alpha_{\mathrm{MLT}}^{{\Delta s}}  / \alpha_{\mathrm{MLT,\sun}}^{{\Delta s}}$ } & \citetalias{Magic2015} & 4.968 $\pm$ 0.271 &  0.076 $\pm$ 0.005  & -1.145 $\pm$ 0.076 & -0.017 $\pm$ 0.003  \\             
\multicolumn{1}{l}{$\alpha / \alpha_{\sun}$ } & \citetalias{Trampedach2014} & 3.174    $\pm$   0.277  & 0.048    $\pm$   0.007 & -0.637   $\pm$    0.076  & N/A \\
\multicolumn{1}{l}{$\alpha_{\mathrm{m}} / \alpha_{\mathrm{m},\odot}$ } & \citetalias{Trampedach2011} & 4.611  $\pm$    0.457 & 0.071  $\pm$    0.010  & -1.037  $\pm$    0.125  & N/A  \\ 
\multicolumn{1}{l}{$s_{\mathrm{bot}} / s_{\mathrm{bot,\sun}}$ } & \citetalias{Magic2015} & -7.413  $\pm$    0.345  & -0.173  $\pm$    0.006  & 2.465    $\pm$  0.096  & 0.051    $\pm$  0.003 \\
\multicolumn{1}{l}{$ \Delta s / \Delta s_{\sun}$} & \citetalias{Magic2015} & -54.586 $\pm$ 3.006 & -1.046 $\pm$ 0.051 & 16.214 $\pm$ 0.835 & 0.288 $\pm$ 0.029 \\
\multicolumn{1}{l}{$ \delta s_{\mathrm{rms}}^{\mathrm{peak}} / \delta s_{\mathrm{rms,\sun}}^{\mathrm{peak}}$} & \citetalias{Magic2015} & -45.571  $\pm$  2.205 & -0.853 $\pm$  0.037 & 13.538 $\pm$  0.613 & 0.221 $\pm$  0.022 \\ 
\multicolumn{1}{l}{$S_{\mathrm{Jump}} / S_{\mathrm{Jump,\sun}}$} & \citetalias{Trampedach2013} & -57.123   $\pm$   6.560  & -1.428   $\pm$   0.159  & 17.299 $\pm$     1.799  & N/A \\
\multicolumn{1}{l}{$S_{\mathrm{max}} / S_{\mathrm{max,\sun}}$} &  \citetalias{Trampedach2013} & -83.366   $\pm$   7.777  & -2.203  $\pm$    0.189  & 25.229  $\pm$   2.132  & N/A  \\   
\multicolumn{1}{l}{$\delta v_{z,\mathrm{rms}}^{\mathrm{peak}} / \delta v_{z,\mathrm{rms,\sun}}^{\mathrm{peak}}$}  & \citetalias{Magic2015} & -20.404 $\pm$    0.623 & -0.410  $\pm$    0.011 &  6.223  $\pm$    0.173  &  0.090  $\pm$    0.006   \\ 
\hline
\end{tabularx}
\end{table}

In addition to  \Alpha,  \cite{Magic2015} tabulate entropy too, which we fit  to Eq.~\ref{eq:trilinear_model}. Specifically, we fit the \cite{Magic2015} values of $\Delta s / \Delta s_{\sun}$ and $s_{\mathrm{bot}} /  s_{\mathrm{bot,\sun}}$, as well as the maximal rms entropy, $\delta s_{\mathrm{rms}}^{\mathrm{peak}}$, to Eq. \ref{eq:trilinear_model}. Additionally, entropy values from \cite{Trampedach2013} (henceforth \citetalias{Trampedach2013}) were obtained. From \cite{Trampedach2013} values of $S_{\mathrm{max}}$, the asymptotic entropy, and $S_{\mathrm{Jump}}$, the atmospheric entropy jump, were obtained. The \cite{Trampedach2013} entropy values were also fit to Eq.~\ref{eq:trilinear_model}. Note that the \cite{Trampedach2013} models are all of solar metallicity, and so the metallicity coefficient cannot be determined for these models. The best-fit coefficients for the entropy values can be found in Table~\ref{table:simulation_compare}. Comparing the best-fit coefficients for the entropy values to the results of this work, it is interesting to note that the metallicity dependence of our $\alpha$ and the metallicity dependence of the \citetalias{Magic2015} entropy measures have the same sign. However, since a larger $\alpha$ in models implies a smaller entropy jump, in essence the results are again in disagreement. It should be noted that mixing-length models ignore the effects of turbulent pressure --- gas pressure alone supports gravity, this of course changes what would have been the entropy and thus comparing entropy may not be a fair comparison, particularly since the presence of turbulence effectively changes the equation of state.

\cite{Magic2015} also had convective velocities. We performed the trilinear fit to the maximal rms velocity, $\delta v_{z,\mathrm{rms}}^{\mathrm{peak}}$, as well. The best-fit coefficients can be seen in Table~\ref{table:simulation_compare}. This is the only case where we see an agreement with the sign of the metallicity dependence. Under the mixing-length approximation, a larger $\alpha$ implies a larger velocity, and thus both convection simulations and mixing length models show larger velocities for larger metallicities. Larger convective velocities for higher metallicity simulations were also seen in the simulations of \cite{Tanner2013}, implying that this is a robust feature of both convection simulations and mixing-length models.

\subsection{Summary and Conclusions}
Stars with observed values of $\nu_{\mathrm{max}}$, $\Delta \nu$, $T_{\mathrm{eff}}$, and \FeH\ were modeled in YREC. The resulting likelihood weighted average stellar properties were compared to the mixing length parameter for each star. We found that for the non-diffusion, $0.2H_p$ set of models, a linear equation of the form $\alpha/\alpha_{\sun}=5.426 -0.101 \log (g)  -1.071 \log (T_{\mathrm{eff}}) + 0.437 (\mathrm{[Fe/H]})$ best represented the relationship between \Alpha\ and the stellar parameters. This process was repeated for several sets of stellar model input physics. The signs of the best-fit coefficients for the linear model  were all found to agree with \cite{Bonaca2012}. The equational form of the relationship between $\log(g)$, $\log(T_{\mathrm{eff}})$, \FeH, and \Alpha\ was also explored using the \textit{Eureqa} symbolic regression software. The results were also compared to values of \Alpha\ determined from 3D convection simulations, however the trends observed in this work did not fully agree with the relationships observed from the simulations. The impact of a metallicity dependent mixing length value was demonstrated through the creation of two sets of example isochrones, of $\mathrm{[Fe/H]}=-0.5 \: \mathrm{and} \: +0.5$, constructed with both the solar mixing length and the metallicity dependent mixing length values from this work. The large effect that the mixing length parameter has on stellar models, combined with what we know about the shortcomings of assuming a solar calibrated mixing length, make understanding the relationship between \Alpha\ and stellar parameters, especially metallicity, vitally important. Further investigation into the disagreement between the results of this work and those from 3D convective simulations, especially the disagreement in the metallicity dependence of \Alpha, is needed.

\acknowledgments This work was partially supported by NSF grant AST-1514676 and NASA grant NNX16AI09G to SB. AB was supported by an Institute for Theory and Computation Fellowship. WJC acknowledges the support of the UK Science and Technology Facilities Council (STFC). Funding for the Stellar Astrophysics Centre is provided by The Danish National Research Foundation (Grant DNRF106).

\software{YREC \citep{Demarque2008}, \textit{Eureqa} \citep{SchmidtLipson2009}}

\bibliography{Mixing_Length.bib}

\end{document}